\documentstyle[prl,aps,preprint,epsf]{revtex} 
\large  

\begin{document}
\draft

\title{Binary data corruption due to a Brownian agent II:\\
two dimensions, competing agents, and generalized couplings} 
\author{Wannapong Triampo\cite{wta} and T. J. Newman\cite{tjna}} 
\address{Department of Physics,\\ 
Virginia Polytechnic Institute and State University,\\ 
Blacksburg, VA 24061, USA} 
\maketitle
\begin{abstract}
This work is a continuation of our previous investigation
of binary data  corruption due to a Brownian agent
[T. J. Newman and W. Triampo, preprint cond-mat/9811237]. We extend our
study in three main directions which allow us to make closer contact with
real bistable systems. These are i) a detailed analysis of two dimensions, 
ii) the case of competing agents, and iii) the cases of asymmetric and
quenched random couplings. 
Most of our results are obtained by extending our original phenomenological 
model, and are supported by extensive numerical simulations. 
\end{abstract}
\vspace{5mm} 

\newpage

\section{Introduction}

This paper is the second part of a two-stage investigation into the 
statistics of an active random walker (Brownian agent) in a bistable medium.
This is but one example of the myriad of systems in which a random
walker interacts in some way with its environment\cite{hug,rwe}.
We consider systems in which the Brownian agent (BA) performs a pure, unbiased
random walk in a medium composed of elements which may take one of two values.
On visiting a given element the BA has a certain probability of switching the
value of that element. 
In our first paper\cite{nt} (hereafter referred to as DCI) we motivated
our investigation into such processes using the example of data corruption
caused by a Brownian agent ({\it i.e.} the elements were taken to be bits
of binary data) and we outlined the possible applications of our results
to describing soft error production in small-scale memory devices
\cite{mid,shar}. The
disturbance of a bistable medium by a BA is also related to reversible 
chemical kinetics by a high mobility catalyst, and disordering of bi-atomic
structures by a wandering agent\cite{vmdr,al}, such as an anion or cation 
vacancy in {\it NaCl}, or an impurity in a semiconductor 
compound ({\it e.g.} {\it Zn} in {\it GaAs}). 
One may also view this process as a non-conserved
cousin of magnetic disordering via spin exchange with a wandering 
vacancy\cite{hh,new,schmitt}. Last, but not least, the analysis of
this process yields a deeper understanding of the statistics of random
walks.

In DCI we studied the simplest possible process, namely
a single BA disordering a bistable system, with a switching probability
which is independent of the value of the element. One of our main
concerns in DCI was to construct a phenomenological continuum model of the
process. This enabled us to find results independent of microscopic details,
and also to study coarse-grained quantities, such as the probability 
distribution for the local ``density of disorder'' which is less easily defined
on a lattice. Our results were derived mostly for the case of one spatial 
dimension. 

In order to make contact with a wider range of processes it is necessary to
consider a more general model. This is the aim of the present work. We shall
extend our original investigation in three directions. First, we shall
present a careful analysis of two spatial dimensions. We find that the
phenomenological model predicts the correct asymptotic behavior, despite the
need to introduce some form of regularization. We present results for
the mean density of disorder ${\bar \rho}$, 
and using some special properties of the
continuum description, we shall also derive an approximate form for the
probability distribution of the density of disorder. Second, we shall
consider a system containing more than one BA, thereby inducing ``competition''
as each BA interferes with the disorder created by the others. The main
result here is that the disordering efficacy 
({\it i.e.} the global amount of disorder due to $N$ agents as compared to 
one agent) is massively reduced
for dimensions less than two, whereas in precisely two dimensions, each BA
eventually becomes independent (in that the disorder it creates is not 
reordered by other BA's). We shall present calculations based on the
continuum model to make these statements quantitative. Third,
we shall consider two kinds of generalized couplings between the BA and the 
bistable medium: asymmetric switching probabilities, and quenched random 
switching probabilities.  We shall argue that these generalized couplings
may be modeled
within the continuum limit by simple generalizations of our original
model, and we shall derive some basic consequences, which for the case
of quenched randomness are particularly interesting (see the outline below). 
In all cases, we shall support our results by numerical simulations
of the underlying lattice model.

A more detailed outline follows: In section II we present a recapitulation
of the results of DCI. We shall briefly describe the lattice model
in general spatial dimension $d$, and
present it in the form of a master equation. We then describe the 
associated continuum theory, and we state without derivation 
some pertinent results 
previously derived from this continuum model in DCI. 
In section III we concentrate on calculating ${\bar \rho}$
in two spatial dimensions. Given
that two is the critical dimension of the process, the results are modulated
by logarithmic corrections. Therefore great care is needed to compare
different theoretical predictions and numerical results. 
We shall present (briefly) four
alternative methods of calculation, and show that they all predict the
same asymptotic behavior, and agree with the numerical simulations once
the strong corrections to scaling are included. Using these results, we
are also able to approximately reconstruct the probability distribution
for the density of disorder. In section IV we study the
case of more than one BA in the context of a generalized version of our 
continuum model. We shall derive an exact integral expression 
(for asymptotically large times)
for the disordering efficacy $\sigma _{N}(d)$ 
for arbitrary $N$. First we concentrate on $d=1$. 
We evaluate $\sigma _{N}(1)$
for $N=2,3,4$, and also extract its functional form for $N \gg 1$. 
We then extend our study to
arbitrary spatial dimension $d < 2$, and evaluate $\sigma _{2}(d)$ and 
$\sigma _{N}(d)$ for large $N$. We use these results to analytically continue 
to two dimensions, thereby avoiding the use of a microscopic regularization.
In section V we consider the system in $d=1$ with generalized couplings.
First we study asymmetric rates, so, for example, in the data corruption
process the BA will have different probabilities to switch $0\rightarrow 1$
and $1\rightarrow 0$. We propose to
model this using a simple extension of the original continuum theory,
based on the idea that relative to the non-zero ``background disorder''
the dynamics of the system are the same as the symmetric case.
Second, we consider quenched random couplings. In this case we argue 
that ${\bar \rho}$ picks up logarithmic corrections in time,  
while the global amount of disorder remains unaffected. We shall also discuss
the quenched average of the distribution function of disorder 
density, and show that it is very sensitive to the distribution of the
couplings. We shall support our results by numerical simulations which are
described in detail in section VI. We end the paper with a 
summary of our results and our conclusions.

\section{Recapitulation}

In this section we give a very brief review of the main ideas and some
of the results contained in DCI in order to place the present work in
a proper context. The process of a BA in a bistable medium is first modeled
on a hypercubic lattice of dimension $d$. 
The position of the BA is denoted by a lattice vector ${\bf R}$.
In a time step $\delta t$ the BA has a probability $p$ to move to one of
its $2d$ nearest neighbor sites. In making such a jump, there is a
probability $q$ that the element on the site departed from is switched.
The elements are described by spin variables ${\sigma }_{\bf r}$ (where 
${\bf r}$ denotes a discrete lattice vector) which may take the values
$\pm 1$. The spin variables encode the information about the disordering
process. For example in the data corruption process we label uncorrupted bits
(of value 1) by spin +1 and corrupted bits (of value 0) by $-1$. 
[Thus, we shall often use the terms
``magnetization density'' and ``global magnetization'' which may be 
simply translated as ``density of disorder'' and ``total amount of 
disorder''.] This process is illustrated in Fig.1 for $d=2$ and $p=q=1$. 
We can define the dynamics via the probability distribution
$P({\bf R},\lbrace \sigma _{\bf r} \rbrace,t)$, which is the probability that
at time $t$, the BA is at position ${\bf R}$ and the spins have values given
by $\lbrace \sigma _{\bf r} \rbrace$. This distribution evolves according
to a master equation\cite{gard} which takes the form
\begin{eqnarray}
\label{master}
\nonumber
P({\bf R}, \lbrace \sigma _{\bf r} \rbrace , t+\delta t)  = 
(1-p) P({\bf R}, \lbrace \sigma _{\bf r} \rbrace , t) & + & {p(1-q)\over 2d}
\sum \limits _{\bf l}P({\bf R}+{\bf l}, \lbrace \sigma _{\bf r} \rbrace , t)\\
& + & {pq\over 2d}
\sum \limits _{\bf l}P({\bf R}+{\bf l}, \cdots, -\sigma _{{\bf R}+{\bf l}},
\cdots , t) \ ,
\end{eqnarray}
where $\lbrace {\bf l} \rbrace$ represents the $2d$ orthogonal lattice vectors
(which have magnitude $l$).

The most direct quantities to extract from the master equation are marginal
averages, the simplest of which is the magnetization density given by
\begin{equation}
\label{lomag}
\Theta ({{\bf r}_{1}}, {\bf R}, t) \equiv {\rm Tr} _{\sigma} \ 
\sigma _{{\bf r}_{1}} \ P({\bf R}, \lbrace \sigma _{\bf r} \rbrace, t) \ .
\end{equation}
Averaging the master equation over the spin variables gives 
\begin{eqnarray}
\label{lomageq}
\nonumber
\Theta ({\bf r}, {\bf R}, t+\delta t )  - 
\Theta ({\bf r}, {\bf R}, t)  & = & {p\over 2d}
\sum \limits _{\bf l} \bigl [ \Theta ({\bf r}, {\bf R}+{\bf l}, t) - 
\Theta ({\bf r}, {\bf R}, t) \bigr ]\\
& - & {pq\over d} \ \Theta ({\bf r}, {\bf r}, t) 
\sum \limits _{\bf l} \delta _{{\bf r}, {\bf R}+{\bf l}} \ .
\end{eqnarray}

At this stage a continuum limit (in both space and time) may be taken of the 
above equation, which yields
\begin{equation}
\label{lomco}
\partial _{t}\Theta ({\bf r}, {\bf R}, t)  =
{D\over 2} \nabla _{\bf R}^{2} \Theta ({\bf r}, {\bf R}, t) 
 - \lambda \ \Theta ({\bf r}, {\bf R}, t) \Delta _{\bf l}({\bf r}-{\bf R}) \ .
\end{equation}
Two parameters have appeared: the effective diffusion constant of the BA
given by $D=2l^{2}p/\delta t$, and an effective coupling between the BA 
and the spins given by $\lambda \propto pql^{d}/\delta t$. 
This continuum equation
for $\Theta$ has the form of a diffusion equation with a sink potential
$\Delta _{\bf l}({\bf r})$ which is a strongly localized function with
lateral extent $l$ and normalized to unity. 
In the naive continuum limit, this function may be taken
to be a $d$-dimensional Dirac delta function. However, for $d \ge 2$ it is
necessary to smear this function in order to regularize
the theory.

In DCI an alternative continuum description was obtained by viewing the 
process as a stochastic cellular automaton. The process is then defined in
terms of the position ${\bf R}(t)$ of the BA (which is now an independent
stochastic process), and the coarse-grained
density of disorder (or magnetization density)
which is defined in a small region of space at a specific time, and is
a functional of ${\bf R}(t)$. In some sense, one may view this 
in the same spirit as a Langevin description of a stochastic process described
at a more fundamental level by a master equation. Taking the continuum limit
of this description yields first a simple Langevin equation for the
position of the BA
\begin{equation}
\label{scacont1}
{d {\bf R} \over dt} = {\bf {\xi }}(t) \ ,
\end{equation}
where ${\bf {\xi}} (t)$ is a noise term, each
component of which is an uncorrelated Gaussian random variable with zero 
mean ({\it i.e.} $\xi _{i} (t)$ is a white noise process). 
The correlator of ${\bf {\xi}}$ is given by 
\begin{equation}
\label{corrxi}
\langle \xi _{i} (t) \xi _{j}(t') \rangle = D'\delta _{i,j}\delta (t-t') \ .
\end{equation}
Here and henceforth, angled brackets indicate an average
over the noise (or equivalently the paths of the BA). The BA is chosen to
reside initially at the origin: ${\bf R}(0)={\bf 0}$.

The evolution of the magnetization density $\phi $ is described by
\begin{equation}
\label{scacont2}
\partial _{t} \phi ({\bf r},t) = -\lambda ' \phi ({\bf r},t)
\Delta _{l} ({\bf r}-{\bf R}(t)) \ .
\end{equation}
This equation may be integrated to give the explicit functional solution
\begin{equation}
\label{solcon}
\phi ({\bf r},t) = \exp \left [ -\lambda ' \int \limits
_{0}^{t} dt' \ \Delta _{l} ({\bf r}-{\bf R}(t')) \right ] \ . 
\end{equation}
The above solution is obtained for an initial condition $\phi ({\bf r},0)=1$,
which we shall use exclusively. In terms of the original lattice model it
corresponds to choosing all the spins to have the initial value of +1,
so that we measure the subsequent disorder of the system by counting the
number of minus spins in the system. [Although of no relevance to the
coarse-grained description, we mention here that it is often convenient to
choose the spin at the origin ({\it i.e.} the initial BA position) to be
$-1$, so that for $p=q=1$ all spins have value +1 after the first jump of the
BA.]

In DCI we showed that the continuum descriptions (\ref{lomco}) and 
({\ref{scacont2}) are indeed equivalent with the identifications $D=D'$ and
$\lambda = \lambda'$. [This was proven by considering the former as an
imaginary time Schr\"odinger equation\cite{ll}, and writing the solution of
the latter as an imaginary time Feynman path integral\cite{fh}.]
This ends our review of the modelization of the process -- the reader is
referred to DCI for further details and discussions.

In DCI we exclusively used the continuum description (\ref{solcon}) to 
generate results for various average quantities. The simplest quantity
to consider is the mean magnetization density given by 
$m({\bf r},t)=\langle \phi ({\bf r},t) \rangle$  (which is equivalent to
the sum of $\Theta ({\bf r},{\bf R},t)$ over the BA position ${\bf R}$).
For $d<2$ there is no necessity for regularization and we replaced the
sink function $\Delta _{l}$ by the $d$-dimensional Dirac delta function.
The (temporal) Laplace transform for $m({\bf r},t)$ was
found to have the exact form
\begin{equation}
\label{ltden}
{\hat m}({\bf r},s) = {1 \over s} \left [ 1 - {\lambda {\hat g}({\bf r},s) 
\over 1+\lambda {\hat g}({\bf 0},s)} \right ] \ ,
\end{equation}
where ${\hat g}({\bf r},s)$ is the Laplace transform of the diffusion equation
Green function
\begin{equation}
\label{greenlt}
g({\bf r},t) = (2\pi Dt)^{-d/2} \exp (-r^{2}/2Dt) \ .
\end{equation}

This exact result allows one to extract a great deal of statistical information
about the process. First, one may simply invert the Laplace transform to
find the average magnetization density (or average
density of disorder relative to
1/2) as a function of ${\bf r}$ and $t$. Explicit forms are given in DCI for 
$d=1$. We note here that the average
magnetization density at the origin decays for long times in $d=1$ as
\begin{equation}
\label{resden0}
m(0,t) = \left ( {2D \over \pi \lambda ^{2} t} \right )^{1/2}
\left [ 1 + O \left ( {D \over \lambda ^{2}t} \right )
\right ] \ . 
\end{equation}
The continuum solution has the important property that
$\langle \phi ({\bf r},t;\lambda)^{n} \rangle = 
\langle \phi ({\bf r},t;n\lambda) \rangle$. This allows us to utilize the
exact solution (\ref{ltden}) to reconstruct the probability density for
the magnetization density. We define ${\cal P}$ via
\begin{equation}
\label{probden}
{\cal P}(\phi, {\bf r}, t) = 
\langle \delta (\phi - \phi _{\bf R}({\bf r},t)) \rangle \ ,
\end{equation}
where $\phi _{\bf R}({\bf r},t)$ is the stochastic field solution given in Eq.
(\ref{solcon}).
As explained in DCI, one may solve for this distribution exactly.
In particular, for $d=1$ 
the probability distribution for the magnetization density
at the origin takes the form 
\begin{equation}
\label{lognor}
{\cal P}(\phi , 0 , t) = {1 \over (\pi t)^{1/2}} \ {1 \over {\hat \lambda}
\phi } \exp \left [ -{(\log (\phi ))^{2} \over 4{\hat \lambda }^{2}t}
\right ] \ ,
\end{equation}
which is a log-normal distribution (and where we have defined
${\hat \lambda}=\lambda/(2D)^{1/2}$). This is interesting,
as it indicates the extreme nature of the fluctuations in this system.
For instance, the typical value of the magnetization density can be found
from the above expression to decay exponentially, {\it i.e.}
$\phi _{\rm typ} \sim \exp (-{\hat \lambda} ^{2}t/2)$, whereas the
mean density decays as $1/{\hat \lambda }\sqrt{t}$ as given in (\ref{resden0}).

Another interesting quantity which may be extracted from $m({\bf 0},t)$
is the average global magnetization defined (relative 
to its initial value) as
\begin{equation}
\label{glomagc1}
M(t) = \int d^{d}r \ \left [
\langle \phi ({\bf r},0) \rangle - \langle \phi ({\bf r},t) 
\rangle \right ] \ .
\end{equation}
As shown in DCI, for $d<2$ this quantity obeys the exact relation
\begin{equation}
\label{glomagc3}
{dM(t) \over dt} = \lambda m({\bf 0},t) \ .
\end{equation} 
Thus, for large times in one dimension 
we have $M(t) \sim (Dt)^{1/2}$ independent of the
coupling $\lambda $. In other words, the total amount of disorder created
by a single BA on average increases as $(Dt)^{1/2}$, and is (rather 
surprisingly) independent of the coupling between the BA and the spins
(for times larger than $D/\lambda ^{2}$).

\section{Two dimensional systems}

In this section we shall present a careful analysis of the case of two
dimensions. The simple random walk is recurrent for dimensions $d \le 2$,
whereas for $d>2$ the walker has a probability less than unity for ever
returning to its starting point\cite{hug}. This basic fact from the theory
of random walks has an immediate implication for our data corruption
problem. The non-recurrent nature of random walks for $d>2$ implies that
the BA will continually corrupt new regions of the system, and rarely
revisit sites which it has previously corrupted. Thus the relative density of
disorder (or average magnetization density) at the origin $m({\bf 0},t)$
will decay to a 
non-zero (and non-universal) value, and the total amount of disorder
(or average global magnetization) $M(t)$ will increase linearly in time, with
a non-universal prefactor. For practical applications, in which one wishes
to limit the disordering capabilities of the BA, the first requirement
is to restrict the geometry of the system to a dimension $d \le 2$.
So $d=2$ is the critical dimension of the problem, and because of this
we can expect logarithmic corrections to modulate the leading order
results, and also to cause long cross-over times, thus making numerical
results more difficult to interpret. 

In DCI we studied some general properties of higher dimensional systems
and we also derived an approximate form for $m({\bf 0},t)$ in $d=2$ 
using a crude
form of regularization. Rough agreement was found between this result and
simulations, but no quantitative data analysis was performed. In this
section we shall rederive the form of $m({\bf 0},t)$ more carefully.
The reasons for this are threefold: first, so that we can have confidence 
in the leading order result, and also get some idea of the sub-leading
corrections; second, to provide insight into the relation between the
discrete and continuum approaches in two dimensions (which is important
given that the latter must be regularized for $d=2$); and third, to allow us
to construct the form of the probability distribution of the magnetization
density (for which we need as much information about $m({\bf 0},t)$ as 
possible). 

We shall use four different methods (which will each be described with
brevity) to derive the form of $m({\bf 0},t)$. Each has its strong and
weak points as we shall see. The methods to be used are i) calculation
from the exact lattice formulation for the marginal average (\ref{lomageq}),
ii) solution of the diffusion equation (\ref{lomco}) 
using a smeared sink function,
iii) calculation with infinite-order perturbation theory of the continuum
theory (\ref{solcon}) using a crude temporal cut-off (as described in DCI), 
and finally iv) analytic continuation from the
exact result (\ref{ltden}) valid for $d<2$.

\subsection{Lattice calculation from Eq.(\ref{lomageq})}

Referring to the equation of motion (\ref{lomageq}) for the marginal 
average we set $d=2$, and for convenience we set the hopping probability 
$p=1$, giving
\begin{equation}
\label{lomageq1}
\Theta ({\bf r}, {\bf R}, t+\delta t ) =
{1\over 4} \sum \limits _{\bf l} \Theta ({\bf r}, {\bf R}+{\bf l}, t)  
- {q\over 2} \ \Theta ({\bf r}, {\bf r}, t) 
\sum \limits _{\bf l} \delta _{{\bf r}, {\bf R}+{\bf l}} \ .
\end{equation}
As an initial condition we take the BA to be located at the origin, and all 
spins to be +1, except the spin at the origin which is taken to be $-1$. 
[This convention is useful for $q=1$ so that all spins have value +1 
after one time step. However if $q\ll 1$ it
is more convenient to take the spin at the origin to be initially +1 as
the chance of it flipping after one time step is small. We stress that these
different choices for the initial value of the spin at the origin have
no effect on the asymptotic properties of the system, and only serve to
smoothen the magnetization density in the immediate vicinity of the origin.] 
Thus
$\Theta ({\bf r},{\bf R},0)=\delta _{{\bf R},{\bf 0}}
(1-2\delta _{{\bf r},{\bf 0}})$. The solution of (\ref{lomageq1}) may be
attained by discrete Fourier and Laplace transform. Defining the former via
\begin{equation}
\label{discft}
{\tilde \Theta}({\bf r},{\bf k},t) = \sum \limits _{\bf R} 
\Theta({\bf r},{\bf R},t) e^{i{\bf k}\cdot {\bf R}}
\end{equation}
and the latter via
\begin{equation}
\label{disclt}
{\hat \Theta}({\bf r},{\bf R},z) = \sum \limits _{n=0}^{\infty} 
\Theta({\bf r},{\bf R},n\delta t) \ z^{n} \ ,
\end{equation}
it is fairly straightforward to diagonalize Eq.(\ref{lomageq1}) to the form
\begin{equation}
\label{diagmag}
{\hat {\tilde \Theta}}({\bf r},{\bf k},z) = {(1-2\delta _{{\bf r},{\bf 0}})
-2qzf({\bf k})e^{i{\bf k}\cdot {\bf r}} \ {\hat \Theta}({\bf r},{\bf r},z)
\over 1 - zf({\bf k})} \ ,
\end{equation}
where $f({\bf k}) = [{\rm cos} \ k_{1}l + {\rm cos} \ k_{2}l ]/2$.
One may now solve the above equation self-consistently for 
${\hat \Theta}({\bf r},{\bf r},z)$ by inverting the discrete Fourier transform.
One has
\begin{equation}
\label{selfc}
{\hat \Theta}({\bf r},{\bf r},z) = {(1-2\delta _{{\bf r},{\bf 0}})
\int d{\bf k} \ e^{-i{\bf k}\cdot {\bf r}} D({\bf k},z) \over
1 + 2qz \int d{\bf k} \ f({\bf k}) D({\bf k},z) } \ ,
\end{equation}
where $D({\bf k}) = [1-zf({\bf k})]^{-1}$ and the momentum integrals are over
the two-dimensional Brillouin zone. The average magnetization density at the
origin is given by summing $\Theta ({\bf 0},{\bf R},t)$ over ${\bf R}$,
which is equivalent to the zero Fourier mode 
${\tilde \Theta } ({\bf 0},{\bf 0},t)$. Thus, substituting (\ref{selfc})
into (\ref{diagmag}) we have after some rearrangement
\begin{equation}
\label{finallt}
\sum \limits _{\bf R} {\hat \Theta }({\bf 0},{\bf R},z)
= {-1 \over (1-z)} \left [ {1 + q' (1-z)\int d{\bf k} \ D({\bf k},z) \over
 1 + q' \int d{\bf k} \ D({\bf k},z)} \right ] \ ,
\end{equation}
where $q'\equiv 2q/(1-2q)$.
[The function $ \int d{\bf k} \ D({\bf k},z) $ is very well known in
the theory of random walks\cite{hug}, and is the discrete Laplace transform
of the probability of a random walker to return to its starting point
after $n$ steps.] Finally we must inverse Laplace transform the above equation.
For large $n$ we can extract the asymptotic form of the average magnetization
by invoking the Tauberian theorem\cite{hug}. In this case we need the form of
the Laplace transform as $z \rightarrow 1$. Using\cite{hug}
\begin{equation}
\label{ingfn}
\int d{\bf k} \ D({\bf k},z) \sim {1 \over \pi} \log \left ( {8 \over 1-z}
\right ) \ [ 1 + O(1-z) ] \ ,
\end{equation}
we have from the Tauberian theorem the asymptotic result
\begin{equation}
\label{finaldis}
\sum \limits _{\bf R} \Theta ({\bf 0},{\bf R},n \delta t)
= {-1 \over (q'/\pi)\log n + C(q) } \ .
\end{equation}
The constant $C(q)$ is not accessible from the Tauberian theorem, although it
could in principle be calculated from a careful inverse Laplace transform
of Eq.(\ref{finallt}).

In our simulations we generally take the switching probability $q$ to be 
unity which implies $q'=-2$ and thus
\begin{equation}
\label{finaldis2}
\sum \limits _{\bf R} \Theta ({\bf 0},{\bf R},n \delta t)
= {1 \over (2/\pi)\log n - C(1)} \ .
\end{equation}
In section VI we shall make a direct comparison of this result with 
numerical simulations in two dimensions.
It is also interesting to note from Eq.(\ref{finallt}) that setting the 
switching probability $q=1/2$ gives 
$\sum \limits _{\bf R} \Theta ({\bf 0},{\bf R},n \delta t) = -\delta _{n,0}$.
In other words, the average value of the spin at the origin remains at zero
after one jump of the agent. This 
``maximal uncertainty'' for $q=1/2$ only holds
at the origin, and spins at other lattice sites will have a positive mean
for all times. As a final
note, if $q\ll 1$ it is more convenient to choose the initial value of
the spin at the origin to be +1 in which case we find from the foregoing
analysis 
\begin{equation}
\label{finaldis3}
\sum \limits _{\bf R} \Theta ({\bf 0},{\bf R},n \delta t)
= {1 \over (2q/\pi)\log n + C(q)} \ .
\end{equation}

\subsection{Diffusion equation (\ref{lomco}) 
with smeared sink function}

We now wish to solve for the coarse-grained magnetization density 
$m({\bf 0},t)$ within some continuum limit. In this subsection we 
accomplish this by solving the diffusion equation (\ref{lomco}) with a smeared
sink function. 
This calculation is the closest in spirit to the lattice calculation
since (\ref{lomco}) is the direct continuum analog of the discrete equation
(\ref{lomageq}) solved in the previous subsection. Setting ${\bf r}=0$,
Eq.(\ref{lomco}) takes the form 
\begin{equation}
\label{lomco2}
\partial _{t}\Theta ({\bf 0}, {\bf R}, t)  =
{D\over 2}\nabla _{\bf R}^{2} \Theta ({\bf 0}, {\bf R}, t) 
 - \lambda \ \Theta ({\bf 0}, {\bf R}, t) \Delta _{l}({\bf R}) \ .
\end{equation}
The simplest finite range form to take for the sink function $\Delta _{l}$ is 
a radial function which is zero outside a radius $l$,
and (through normalization) equal to $(1/\pi l^{2})$ within this radius.
It is also convenient to smear the initial condition of $\Theta $ in a
similar way:
\begin{eqnarray}
\label{initc}
\Theta ({\bf 0},{\bf R},0) = \left \lbrace 
\begin{array}{l}1/\pi l^{2} \ , \ \ |{\bf R}| < l \\
0 \ , \ \ \ \ \ \ \ |{\bf R}| > l \ .
\end{array} \right. 
\end{eqnarray} 
Given the radial symmetry of both the sink function and the initial condition,
$\Theta ({\bf 0},{\bf R},t)$ is a function only of the radial coordinate
$R$ and $t$. Thus we define $\Theta _{1}(R,t)$ and $\Theta _{2}(R,t)$
to describe the original function respectively inside and outside the radius 
$l$. On Laplace transforming these functions in time, it is straightforward
to show using boundary value techniques that 
\begin{eqnarray}
\label{bvtec}
\nonumber
{\hat \Theta} _{1}(R,s) & = & {1 \over \pi l^{2} (s + \lambda ')}
+ A(s) I_{0}\left ( \sqrt {s+ \lambda '}  \ R' \right ) \\
{\hat \Theta} _{2}(R,s) & = & 
B(s) K_{0}\left ( \sqrt {s}  \ R' \right ) \ ,
\end{eqnarray}
where we have introduced the scaled quantities $R'=R(2/D)^{1/2}$ and
$\lambda ' = \lambda /\pi l^{2}$, and where $I_{0}(z)$ and $K_{0}(z)$ are
modified Bessel functions\cite{as}.

The functions $A(s)$ and $B(s)$ can be determined by demanding continuity
of $\Theta $ and its radial derivative at the boundary $R=l$. Their forms are
complicated and we shall not write them explicitly here.
The average magnetization density $m({\bf 0},t)$ is obtained by 
integrating $\Theta ({\bf 0},{\bf R},t)$ over ${\bf R}$. Using the properties
of integrals of modified Bessel functions\cite{as} we have the closed form
expression
\begin{equation}
\label{ltcle}
{\hat m}({\bf 0},s) = {1 \over s + \lambda '} +
{2\lambda ' \zeta (s) \over l's^{1/2}(s+\lambda ')^{3/2}} \ ,
\end{equation}
where 
\begin{equation}
\label{ltcleaux}
\zeta (s) =  {K_{1}(\sqrt{s} \ l') \ I_{1}(\sqrt{s+\lambda'} \ l') \over
\sqrt {s} K_{1}(\sqrt{s} \ l')I_{0}(\sqrt{s+\lambda'} \ l')
+ \sqrt {s+\lambda'} \ K_{0}(\sqrt{s} \ l')I_{1}(\sqrt{s+\lambda'} \ l') } \ ,
\end{equation}
and $l'=l(2/D)^{1/2}$.
We may now take the lateral size $l$ of the sink function to zero, and retain
the leading terms in the above expression. Using the properties of Bessel
functions at small argument\cite{as} we finally arrive at
\begin{equation}
\label{ltcle2}
{\hat m}({\bf 0},s) = {1 \over s} \left [ {1 \over w({\tilde \lambda} ) - 
{\tilde \lambda}\log (sl'^{2}) } \right ] \ ,
\end{equation}
where ${\tilde \lambda } = {\lambda / 2\pi D}$ and $w(x) = \sqrt{x}
I_{0}(2\sqrt{x})/I_{1}(2\sqrt {x})$ which approaches unity for small $x$.
It is important to mention that this expression is not valid for 
${\rm Re} \ (s) > 1/l'$,
and thus the apparent pole is an artifact of the limit $l \rightarrow 0$.
Examination of the complete form (\ref{ltcle}) indicates there is no pole for
${\rm Re} \ (s) > 0$, as expected on physical grounds.
We refer the reader to Appendix A in which this Laplace transform is inverted.
For large $t$ the result takes the form
\begin{equation}
\label{ltcle3}
m({\bf 0},t) = {1 \over w({\tilde \lambda }) + {\tilde \lambda }\log (t/\tau) }
\ ,
\end{equation}
where $\tau = l'^{2}e^{-\gamma }$ and $\gamma =0.57721...$ is 
Euler's constant. 

Thus we have arrived at our result, and indeed the leading order behavior
has the same functional form as Eq.(\ref{finaldis3}) 
derived in the previous subsection. A closer comparison of the leading
terms indicates that $\lambda \propto q$ for $q\ll 1$ 
as expected on physical grounds. In the present continuum
calculation, we have also extracted an explicit form for the sub-leading term
({\it i.e.} $w({\tilde \lambda })$) which is non-universal and depends
on the precise form  of the smeared sink function.

\subsection{Infinite order perturbation theory of Eq.(\ref{solcon})}

In DCI we concentrated on analyzing the continuum solution Eq.(\ref{solcon})
using infinite order perturbation theory in $\lambda $,
using a Dirac delta function in place of the sink function $\Delta _{l}$. 
For $d<2$ each
term may be averaged, and the series may then be resummed exactly after
Laplace transform. No divergences appear, and one arrives at the exact
result (\ref{ltden}). Exactly in two dimensions a similar procedure may
be used, but now each term in the perturbation expansion diverges after
averaging over the noise. A controlled analysis requires one to work
with a particular sink function, but as noted in DCI, resummation of the
perturbation series is very difficult in this case. Instead, we persisted
in using the Dirac delta function as a sink function, and regularized each
term in the perturbation theory by introducing a short-time cut-off
$t_{0}$, the physical origin of which is the microscopic correlation 
time of the noise. This regularization scheme is not systematic, and this
is one of the reasons for the present more careful analysis. We shall
enter into no details here of the perturbation theory method as a 
comprehensive description may be found in DCI. We also refer the
reader to Ref.\cite{new} where the same regularization scheme was used
in calculations on vacancy-mediated diffusion, and produced scaling
results in accord with exact lattice calculations\cite{hh}.

Using the time cut-off regularization scheme allows one to extract the
dominant contribution from each term in the perturbation expansion, and
to resum the series. The result is
\begin{equation}
\label{cutoffm}
m({\bf 0},t) = {1 \over 1 + {\tilde \lambda }\log (t/t_{0})} \ .
\end{equation}
It is interesting to compare this result with Eq.(\ref{ltcle3})
obtained in the previous 
subsection. They are seen to agree (for small $\lambda $,
in which case $w({\tilde \lambda }) \approx 1$) if we make the 
identification $t_{0} = \tau \ ( = 2l^{2}e^{-\gamma }/D)$. This indeed
supports the role of $t_{0}$ as a microscopic correlation time of the noise,
since it is seen to correspond to the time taken for the diffusion process
to correlate a microscopic region of size $\sim l^{2}$.

\subsection{Analytic continuation from Eq.(\ref{ltden})}

As a fourth method of extracting the form of $m({\bf 0},t)$ for $d=2$ 
we briefly mention analytic continuation from $d<2$. As discussed in
the previous section, no regularization is required in the perturbation
expansion method for $d<2$, and one retrieves the exact result
\begin{equation}
\label{ltdenzero}
{\hat m}({\bf 0},s) = {1 \over s} \left [ {1 \over
1+\lambda {\hat g}({\bf 0},s)} \right ] \ ,
\end{equation}
where
\begin{equation}
\label{grfnz}
{\hat g}({\bf 0},s) = {\Gamma (1-d/2) \over (2\pi D)^{d/2}s^{1-d/2}} \ ,
\end{equation}
and $\Gamma (z)$ is the gamma function\cite{as}. Using the result
$\Gamma (z) \sim 1/z$ for $z \rightarrow 0$ we have for $\epsilon
\equiv 2-d \rightarrow 0$
\begin{equation}
\label{ltdenzero2}
{\hat m}({\bf 0},s) = {1 \over s} \left [ {1 \over
1 - {\tilde \lambda} \log (s/s_{0})} \right ] \ ,
\end{equation}
where $s_{0} = e^{-2/\epsilon }$. 

On comparing this result with Eq.(\ref{ltcle2}) obtained
in section IIIB, we see that they are equivalent (for small $\lambda $,
in which case $w({\tilde \lambda }) \approx 1$) if we make the identification
$1/\epsilon = \log (1/l')$. This relation has only a formal meaning, as
there is no physical sense in continuing dimensionality. However, it is
instructive to learn that the leading order result may be retrieved intact
from analytic continuation. Indeed, we shall use this tool in section IV to
study competing agents in two dimensions.

\subsection{Calculation of the probability distribution}

To complete this section, we shall briefly discuss the probability distribution
of the magnetization density at the origin. As mentioned in section II, the
continuum theory (\ref{solcon}) has the property 
$\langle \phi ({\bf 0},t;\lambda)^{n} \rangle = 
\langle \phi ({\bf 0},t;n\lambda) \rangle$. Thus, knowledge of the $\lambda $
dependence of the first moment allows one to reconstruct all the moments
of the magnetization density, and thus the probability density 
${\cal P}(\phi , {\bf 0},t)$ 
for this quantity. The method for retrieving ${\cal P}$ from the moments
is given in detail in DCI for the case of $d=1$. A similar calculation
suffices for $d=2$, so long as one has an ``accurate'' form for the
mean magnetization density. In the present section we have attacked this
problem from four different directions and have arrived at agreement for the
asymptotically dominant term for $m({\bf 0},t)$. However, as noted in DCI, it
is necessary to have more information than this in order to construct 
${\cal P}$ correctly. If one tries to calculate ${\cal P}$ using 
$m({\bf 0},t;\lambda) \sim 1/{\tilde \lambda }\log t$ one obtains a 
distribution ${\cal P} \sim m({\bf 0},t)/\phi $ which is singular at 
$\phi = 0$. The sub-leading correction to $m({\bf 0},t;\lambda)$ is crucial
to determine the distribution correctly. Our results from the previous
four subsections are in agreement with regard to the time dependence,
but differ in the way in which $\lambda $ appears in the sub-leading term.
We prefer Eq.(\ref{ltcle3}) in that an explicit form $w({\tilde \lambda })$
appears. However, we have been unable so far to reconstruct ${\cal P}$ using 
this form due to the complicated nature of $w$. For small $\lambda $, 
$w({\tilde \lambda }) \approx 1$, and Eq.(\ref{ltcle3}) then coincides with
the less controlled results (\ref{cutoffm}) and (\ref{ltdenzero2}) 
of subsections IIIC and IIID respectively. The reconstruction of ${\cal P}$
is possible from these forms and one finds
\begin{equation}
\label{prob2}
{\cal P}(\phi, {\bf 0},t) = \beta(t) \phi ^{\beta (t)-1} \ ,
\end{equation}
with $\beta (t) = 1 / {\tilde \lambda }\log (t/t_{0})$. As $t \rightarrow
\infty $ this distribution approaches the form $m({\bf 0},t)/\phi $, but
is never singular for finite times. In section VI we shall describe our
attempts to measure ${\cal P}$ for two dimensional systems. Our numerical
results are in surprisingly good agreement with Eq.(\ref{prob2}) above.

\section{Competing agents}

In this section we shall analyze the effects of many BA's within the system.
We shall assume the BA's to be non-interacting, in the sense that they
are unaware of each other's immediate presence. The non-trivial statistics
reside in the fact that the disordering effects of the BA's statistically 
interact via the overlap of the BA histories. As
we have already seen, a single BA interferes with the previous disorder
it has created, such that the amount of disorder does not simply increase
linearly in time. This effect is more severe when more than one BA is present,
as each BA can disturb the disorder that another BA has previously created.

We measure the strength of this interference via a quantity called the 
``disordering efficacy'' of the agents, defined as
\begin{equation}
\label{def}
\sigma _{N}(d) \equiv \lim \limits _{t \rightarrow \infty}
{M^{(N)}(t) \over M^{(1)}(t) } \ ,
\end{equation}
where $M^{N}(t)$ is the average global disorder created by $N$ agents. If
the BA's were truly independent (in terms of the disorder they create) then
we would expect $\sigma _{N} = N$. As we shall see, for $d<2$ the value of
$\sigma _{N}$ is strongly reduced below this value. However, for $d=2$ this
value is recovered, but only in the deep asymptotic regime ($t \gg e^{N}$).

The extension to many BA's is easily modeled within the continuum theory.
We introduce $N$ random walkers, each of which is described by a position
vector ${\bf R}_{\alpha}(t)$, $\alpha=1,2,\cdots,N$. Since the BA's are 
independent we have
\begin{equation}
\label{scacont1m}
{d {\bf R}_{\alpha} \over dt} = {\bf {\xi }}_{\alpha}(t) \ ,
\end{equation}
where ${\bf {\xi}}_{\alpha} (t)$ are independent Gaussian white noise 
sources with zero mean. The equation of motion for the coarse-grained
magnetization density $\phi ^{(N)}$ takes the form
\begin{equation}
\label{scacont2m}
\partial _{t} \phi ^{(N)} ({\bf r},t) = -\lambda \phi ^{(N)}({\bf r},t)
\sum \limits _{\alpha =1}^{N} \Delta _{l} ({\bf r}-{\bf R}_{\alpha }(t)) \ ,
\end{equation}
with solution
\begin{equation}
\label{solconm}
\phi ^{(N)} ({\bf r},t) = \prod \limits _{\alpha =1}^{N} \exp \left [ -\lambda 
\int \limits
_{0}^{t} dt' \ \Delta _{l}({\bf r}-{\bf R}_{\alpha }(t')) \right ] \ . 
\end{equation}
On averaging over the paths of the $N$ agents, we have the particularly 
simple result
\begin{equation}
\label{avermagm}
m^{(N)}({\bf r},t) \equiv \langle \phi ^{(N)} ({\bf r},t) \rangle
= m^{(1)}({\bf r},t)^{N} \ .
\end{equation}
Thus the average global magnetization is given by
\begin{equation}
\label{glomagm}
M^{(N)}({\bf r},t) \equiv \int d^{d}r \ \left [ m^{(N)}({\bf r},0) - 
m^{(N)}({\bf r},t) \right ] = \int d^{d}r \ \left [ 
1-m^{(1)}({\bf r},t)^{N} \right ] \ .
\end{equation}
Scaling the spatial coordinate by the diffusion length scale $(2Dt)^{d/2}$,
we have from Eqs. (\ref{def}) and (\ref{glomagm})
\begin{equation}
\label{def2}
\sigma _{N}(d) = \lim \limits _{t \rightarrow \infty}
{ \int d^{d}r \ \left [1-m^{(1)}({\bf r}/(2Dt)^{1/2},t)^{N} \right ]
\over \int d^{d}r \ \left [1-m^{(1)}({\bf r}/(2Dt)^{1/2},t) \right ] } \ .
\end{equation}

To proceed with the calculation it is convenient to first perform the 
large-$t$ limit. We concentrate on $d<2$, and therefore replace $\Delta
_{l}({\bf r})$ by the Dirac delta-function.
From Eqs. (\ref{ltden}) and (\ref{greenlt}) it is a fairly 
straightforward matter to show that
\begin{equation}
\label{limitmag}
m(r) \equiv \lim \limits _{t\rightarrow \infty} m^{(1)}({\bf r}/(2Dt)^{1/2},t)
={2 \over \Gamma (1-d/2)} \int \limits _{0}^{r} du \ u^{1-d} \ e^{-u^{2}} \ .
\end{equation}
We may use this result to explicitly evaluate the denominator of 
Eq. (\ref{def2}) giving us
\begin{equation}
\label{def3}
\sigma _{N}(d) = d \ \Gamma (1-d/2) \int \limits _{0}^{\infty} dr \ r^{d-1} \
(1-m(r)^{N}) \ .
\end{equation}
We refer the reader to Appendix B where it is demonstrated that the
above expression may be recast in the more useful form
\begin{equation}
\label{def4}
\sigma _{N}(d) = {2N(N-1)\over \Gamma (1-d/2)} \int \limits _{0}^{\infty}
dr \ r^{1-d} \ e^{-2r^{2}} \ m(r)^{N-2} \ ,
\end{equation}
for $N>1$. We note that the result for two agents follows immediately from
this expression, and we have $\sigma _{2} = 2^{d/2}$. 
This result is striking. For the case of $d=1$, we see that two agents create
only $\sqrt{2}$ as much disorder as one agent. Also, assuming we may continue
this result to exactly two dimensions, we find that $\sigma _{2}(2)=2$ --
{\it i.e.} two agents create disorder independently. In the remainder of this
section we shall use Eq.(\ref{def4}) for two purposes. First, we shall 
concentrate on $d=1$ and evaluate the exact values of $\sigma _{3}(1)$ and
$\sigma _{4}(1)$, which may be used to compare with numerical simulations.
Second, we shall evaluate the integral for large $N$ for arbitrary 
$d \in [0,2]$ using a saddle point method. This calculation will make clear
the tremendous difference in the large-$N$ behavior of $\sigma _{N}(d)$
for $d<2$ and $d=2$. We end the section with a simple scaling argument
which helps us to understand these analytic results.

For $d=1$ the expression (\ref{limitmag}) is simply $m(x)={\rm erf}(x)$,
where ${\rm erf}(z)$ is the error function\cite{as}. Thus we have from
Eq.(\ref{def4})
\begin{equation}
\label{def4_1}
\sigma _{N}(1) = {2N(N-1)\over \pi ^{1/2}} \int \limits _{0}^{\infty}
dx \ e^{-2x^{2}} \ {\rm erf}(x) ^{N-2} \ .
\end{equation}
We are able to evaluate these integrals for $N=3$ and $N=4$. The details
may be found in Appendix C. The results are
\begin{equation}
\label{def5_1}
\sigma _{2}(1) = \sqrt{2} \ , \hspace{1.0cm} \ \sigma _{3}(1) = {6\sqrt{2} 
\over \pi} \ {\rm sin}^{-1} \left ( {1\over
\sqrt {3} } \right ) \ , \hspace{1.0cm} \ 
\sigma _{4}(1) = {12\sqrt{2} \over \pi} \ {\rm sin}^{-1} \left ( {1\over 3 } 
\right ) \ .
\end{equation}
The numerical values of these expressions are presented in Table I along 
with the results from our computer simulations (the details of which may be 
found in section VI). Excellent agreement is found.

\begin{center}
\begin{tabular}{|c||c|c|}\hline
{ \ \ $N$ \ \ } & { \ \ $\sigma _{N}(1)_{\rm theory}$ \ \ } & { \ \ 
$\sigma _{N}(1)_{\rm simul}$ \ \ } \\ \hline 
  2 & $1.414 \cdots$ & $1.42(1)$ \\ \hline
  3 & $1.662 \cdots$ & $1.66(1)$ \\ \hline
  4 & $1.835 \cdots$ & $1.84(2)$ \\ \hline
\end{tabular}
\end{center}

\vspace{0.2cm}

\noindent
Table 1: The predicted values of $\sigma _{N}(1)$ for $N=2,3$ and 4, from
Eq.(\ref{def5_1}), compared with our numerical simulations.

\vspace{0.2cm}

We now return to the case of arbitrary $d \in [0,2]$ and consider the limit
of large $N$. For the sake of generality, we study the integral
\begin{equation}
\label{largenint}
Q_{N}(\beta ,d) = \int \limits _{0}^{\infty} dr \ r^{1-d} \ e^{-\beta r^{2}} \
m(r)^{N} \ ,
\end{equation}
with $m(r)$ as given above in Eq.(\ref{limitmag}). We can recover the 
disordering efficacy via 
\begin{equation}
\label{recov}
\sigma _{N}(d) = {2N(N-1) \over \Gamma (1-d/2)} Q_{N-2}(2,d) \ .
\end{equation}
We wish to implement a saddle point calculation, so we rewrite 
(\ref{largenint}) as 
\begin{equation}
\label{largenint2}
Q_{N}(\beta ,d) = \int \limits _{0}^{\infty} dr \ r^{1-d} \ e^{-F_{N}(r)} \ ,
\end{equation}
where
\begin{equation}
\label{largenint3}
F_{N}(r; \beta ,d) = \beta r^{2} - N\log (m(r)) \ .
\end{equation}
The saddle point is defined via $dF_{N}/dr |_{r=r_{0}} = 0$, which yields
for $r_{0}$ the transcendental equation
\begin{equation}
\label{sadpt}
\beta r_{0}^{d} =  {Ne^{-r_{0}^{2}} \over m(r_{0})\Gamma (1-d/2)} \ .
\end{equation}
For $N \gg 1$, it is easy to see that $r_{0} \sim (\log N)^{1/2} \gg 1$,
so that $F(r_{0}) \gg 1$ and the saddle point method is self-consistently
justified.
There are two stages to the calculation. First we must evaluate $F(r_{0})$
to the desired precision. Second, we must Taylor expand $F(r)$ about the
saddle point to account for fluctuations. [It turns out that we need to
evaluate the leading {\it and} sub-leading terms in order to have a useful
comparison to the numerical data. Thus the calculation is somewhat involved.]
For the first stage we iteratively solve Eq.(\ref{sadpt}) to obtain
\begin{equation}
\label{sadpt2}
r_{0}^{2} = \log \left [ {N \over \beta \Gamma (d/2) (\log N)^{d/2}}
\right ] + {d^{2} \over 4} {\log (\log N) \over \log N} + O\left (
{1 \over \log N } \right ) \ .
\end{equation}
Substituting this result into Eq.(\ref{largenint3}), 
we have after some manipulations
\begin{equation}
\label{sadpt3}
\exp [-F_{N}(r_{0})] = \left [ {\beta \Gamma (1-d/2) (\log N)^{d/2} \over
eN } \right ] ^{\beta } \ \left [ 1 - {\beta d^{2} \over 4} {\log (\log N)
\over \log N } +  O\left ({1 \over \log N } \right ) \right ]  \ .
\end{equation}

The second stage proceeds as follows. We Taylor expand the function $F_{N}$
about its value at the saddle point:
\begin{equation}
\label{taylor}
F_{N}(r) = F_{N}(r_{0}) + \sum \limits _{n=2}^{\infty} {(r-r_{0})^{n} \over
n!} {d^{n} F(r) \over dr^{n}} \Biggl | _{r=r_{0}} \ .
\end{equation}
When using the saddle point method, it is usually only necessary to 
consider the second derivative in the above expansion (the so-called
Gaussian fluctuations). In the present case, it turns out that {\it all} terms
in the expansion contribute equally to the fluctuations. It is relatively
straightforward to show by repeated differentiation of $F_{N}(r)$ that
\begin{equation}
\label{difffn}
{d^{n} F(r) \over dr^{n}} \Biggl | _{r=r_{0}} = \beta (-2r_{0})^{n}
\ \left [ 1 + O\left ({1 \over r_{0}^{2} } \right ) \right ] \ .
\end{equation}
Setting $r=r_{0}+{\hat r}$ we have from Eqs. (\ref{taylor}) and (\ref{difffn})
\begin{equation}
\label{taylor2}
F_{N}(r) = F_{N}(r_{0}) + \beta \sum \limits _{n=2}^{\infty} 
{(-2r_{0} {\hat r})^{n} \over n!} \ \left [ 1 + O\left ({1 \over r_{0}^{2} } 
\right ) \right ] \ .
\end{equation}
We now substitute this expansion into Eq.(\ref{largenint2}). 
Scaling ${\hat r}$ by $2r_{0}$ and explicitly performing the sum over $n$ 
we have
\begin{equation}
\label{largenint4}
Q_{N}(\beta ,d) = {\exp [-F_{N}(r_{0})]\over 2r_{0}^{d}} \int \limits
_{-2r_{0}^{2}}^{\infty} d{\hat r} \ \exp \left [ -\beta \left (
e^{-{\hat r}}+{\hat r}-1 \right ) \right ]  \ \left [ 1 + O\left 
({1 \over r_{0}^{2} } \right ) \right ] \ .
\end{equation}
Performing the integral and neglecting exponentially small terms we have
\begin{equation}
\label{largenint5}
Q_{N}(\beta ,d) = \left ( {e \over \beta } \right )^{\beta} \Gamma (\beta )
\ {\exp [-F_{N}(r_{0})]\over 2r_{0}^{d}}  \ \left [ 1 + O\left 
({1 \over r_{0}^{2} } \right ) \right ] \ .
\end{equation}
Finally, we substitute into this expression the form (\ref{sadpt2}) for
$r_{0}$, and the result (\ref{sadpt3}) for $F_{N}(r_{0})$ to yield
\begin{equation}
\label{largenint6}
Q_{N}(\beta ,d) = {\Gamma (\beta )\Gamma (1-d/2)^{\beta } \over 2}
\ {(\log N)^{(\beta -1)d/2} \over N^{\beta }} 
 \ \left [ 1 - {(\beta -1)d^{2} \over 4} {\log (\log N)
\over \log N } +  O\left ({1 \over \log N } \right ) \right ]  \ .
\end{equation}
Combining Eqs. (\ref{recov}) and (\ref{largenint6}) then gives us the 
final large-$N$ result
\begin{equation}
\label{deflargen}
\sigma _{N}(d) = \Gamma (1-d/2) \ (\log N)^{d/2}  
 \ \left [ 1 - {d^{2} \over 4} {\log (\log N)
\over \log N } +  O\left ({1 \over \log N } \right ) \right ]  \ .
\end{equation}
In particular, for $d=1$ the disordering efficacy increases as 
$(\log N)^{1/2}$ with strong logarithmic corrections. We also note that
for real-valued $d<2$, the disordering efficacy increases as $(\log N)^{d/2}$.
In other words, the BA's overlap very strongly in their disordering for all
$d<2$.

The main reason for calculating $\sigma _{N}(d)$ for real-valued $d<2$ was
to attempt to analytically continue the result to $d=2$ which is otherwise
difficult to calculate due to the presence of an explicit regulator.
However, if we simply set $d=2$ in our expression for $\sigma _{N}(d)$
above, we find that the apparent behavior $\sigma _{N}(2) \sim \log N$
is multiplied by the infinite constant. This gives us the hint
that the large-$N$ behavior for $d=2$ is stronger than $\log N$, but offers
no more information than that. We can, however, trace the failure of the
saddle point calculation for $d=2$ back to the saddle point equation 
(\ref{sadpt}) for $r_{0}$. For large-$N$ and $d<2$ we have the leading
result $r_{0} \sim [\log \epsilon N ]^{1/2}$, where
$\epsilon = 2-d$ (and we have obtained this result from (\ref{sadpt}) 
by taking the
small $\epsilon $ and large $r$ limit of $m(r)$). Thus
for fixed $\epsilon > 0$ we may always take $N$ large enough to create a
saddle-point at a large value of $r_{0}$, in which case the result 
(\ref{deflargen}) is valid. However for fixed $N$ (no matter how large),
taking $\epsilon \rightarrow 0$ squeezes the saddle point back into the
origin, in which case the saddle point method is of no use. 

Therefore we cannot use Eq.(\ref{deflargen}) to analytically continue to
$d=2$. However, a simpler method may be used to extract the result.
We refer the reader to the original expressions for $\sigma _{N}(d)$ and
$m(r)$ as given in Eqs.(\ref{def4}) and (\ref{limitmag}) respectively.
Taking $\epsilon \ll 1$ and $N \gg 1$ these equations take the form
\begin{equation}
\label{def4eps}
\sigma _{N}(d) = \epsilon N^{2} \int \limits _{0}^{\infty}
dr \ r^{\epsilon-1} \ e^{-2r^{2}} \ m(r)^{N} \ ,
\end{equation}
and
\begin{equation}
\label{limitmageps}
m(r) = \epsilon \int \limits _{0}^{r} du \ u^{\epsilon - 1} \ e^{-u^{2}} \ . 
\end{equation}
On studying Eq.(\ref{def4eps}) we see that as $\epsilon \rightarrow 0$ the
integral is dominated by small $r$. Thus we require the small-$r$ form
for $m(r)$ which is easily extracted from (\ref{limitmageps}) to be
$m(r) \simeq r^{\epsilon}$. We now break the integral in (\ref{def4eps}) into
two pieces. The first piece encompasses the range $r \in (0,1)$ so that
we can neglect the Gaussian factor and substitute the small 
$r$ form of $m(r)$ to find the leading order result
\begin{equation}
\label{def4eps2}
\sigma ^{I}_{N}(d) = \epsilon N^{2} \int \limits _{0}^{1} 
dr \ r^{N\epsilon-1}  = N \ .
\end{equation}
The second piece encompasses the range $r \in (1,\infty)$ and for large
enough $N$ will have the asymptotic form $\sigma ^{II}_{N}(d) \sim 
(\log N)/\epsilon $. To summarize, the contribution $\sigma ^{I}_{N}(d)$
dominates for $N \ll 1/\epsilon$, and the contribution $\sigma ^{II}_{N}(d)$
dominates for $N \gg 1/\epsilon$. Thus for $d=2$ we have the result
$\sigma _{N}(2) = N$ for all $N$. This is consistent with the analytic
continuation to $d=2$ of the exact result $\sigma _{2}(d) = 2^{d/2}$ found 
earlier. 

These results may be understood in the following way. Consider $N$ agents
in a system of $d$ dimensions. Since each agent performs a random walk,
the amount of available space in which we can expect to find the agents
has a volume $\sim t^{d/2}$. Also, we know from our previous results that
the amount of disorder created by a single agent increases as $t^{d/2}$
for $d<2$, but as $t/\log t$ for $d=2$. Thus for $d<2$ the available space
for $N$ agents
and the amount of disorder created by a single agent both scale as $t^{d/2}$
which means that there is certain to be interference between the agents,
which will become severe for large $N$ as we have seen. However for $d=2$
the amount of available space for $N$ agents
scales logarithmically faster with $t$ than
does the amount of disorder created by a single agent. Thus for 
times large enough such that $N \ll \log t$ we expect the disorder
created by the agents to become statistically independent and thus
$\sigma _{N}(2)=N$. Note that the ``independence time'' grows exponentially
with the number of agents, making numerical observation difficult for even
moderate values of $N$. Similar arguments have been made in the context
of ``the number of distinct sites visited by $N$ random walkers''\cite{lth}.
In section VI we shall present results for $N=2,3$ and 4 in two dimensions.
The disordering efficacy is seen to slowly approach its expected value.

\section{Generalized couplings}

In all of our work so far we have taken the coupling between the BA and the
spins to be symmetric and spatially homogeneous. At least one
of these properties
is likely to be absent in a practical application of our model. Asymmetry
in the switching probability is likely in data corruption since the states
of a bit are not physically encoded in a symmetric way. Also in chemical
kinetic applications, the reaction rates between two chemical species are
unlikely to be symmetric. Spatial homogeneity of the couplings is also an 
idealized situation. Generally there will be quenched random fluctuations
in the switching rates of different sites, and it is important to quantify
the effect of this randomness on the results obtained so far. In the next
two subsections we shall consider these important generalizations in turn.

\subsection{Asymmetric rates}

Consider the lattice model described in section II with the additional 
property that the probability of flipping a spin depends on the value
of the spin. If the BA leaves a site with spin +1 we flip that spin with
probability $q^{+}$, whereas if the BA leaves a site with spin $-1$ we flip 
that spin with probability $q^{-}$. At a microscopic level the model is
now considerably more complicated as the transition rates for flipping 
depend explicitly on the values of the spins. We shall not construct
the master equation for this case. Rather, we shall try to construct an
analog of the continuum model (\ref{scacont2}) using a simple physical idea.

The main effect of the asymmetric couplings is to favor one type of spin
over the other. For the sake of argument let us take $q^{+}<q^{-}$. In this
case the probability that a site with spin +1 is flipped by the BA is less
than that for a site with spin $-1$, so that after some time there will
be considerably more sites {\it in the active zone} with spin $+1$ than sites 
with spin $-1$. By ``active zone'' we mean the region which has been
thoroughly explored by the BA after an elapsed time $t$. 
This region will have a linear size which 
grows as $\sqrt {Dt}$. We may immediately obtain a rough estimate of the
average global magnetization from this picture. Let us make the crude 
approximation that outside this region the magnetization density 
relative to its initial value of unity is zero, and that inside this region
the relative density is $1-m_{\rm eq}$. The quantity $m_{\rm eq}$ is
the equilibrium value of the average spin value which has the form
\begin{equation}
\label{mequ}
m_{\rm eq} = {q^{-}-q^{+} \over q^{-}+q^{+}} \ .
\end{equation}
We may therefore estimate that the average total magnetization
with asymmetric rates is related to that with symmetric rates via
\begin{equation}
\label{mequ2}
M_{A}(t) = (1-m_{\rm eq}) M_{S}(t) = 
{2 \over (1+q^{-}/q^{+})} \  M_{S}(t) \ ,
\end{equation}
where the subscripts $A$ and $S$ indicate ``asymmetric'' and ``symmetric''
respectively.
[Using this simple argument for the case $q^{+}=q^{-}$ would imply that 
$M(t) \sim (Dt)^{1/2}$, independent of the rates since the quantity 
$m_{\rm eq}=0$. This is indeed the case as described in section II.]

We now make the approximation that in a coarse-grained model the dynamics of 
relaxation to this ``background magnetization'' are the same as the relaxation
to zero magnetization in the symmetric model. This amounts to a simple
linear shift in the order parameter $\phi ({\bf r},t)$ which appears in the
continuum description. We therefore write the analog of (\ref{scacont2})
for asymmetric rates in the form
\begin{equation}
\label{contasy}
\partial _{t} \phi ({\bf r},t) = -\lambda [\phi ({\bf r},t)-m_{\rm eq}]
\Delta _{l} ({\bf r}-{\bf R}(t)) \ .
\end{equation}
We may now use this model to derive results for the average magnetization
density and the average global magnetization. Define the field
${\delta \phi }({\bf r},t) \equiv \phi ({\bf r},t)-m_{\rm eq}$. Then 
${\delta \phi }({\bf r},t)$ satisfies the original continuum theory
(\ref{scacont2}), but with an initial condition $1-m_{\rm eq}$. Thus, for any
realization of the BA we have the relationship
\begin{equation}
\label{relsyasy}
\phi _{A}({\bf r},t;1) - m_{\rm eq} = 
\phi _{S}({\bf r},t;1-m_{\rm eq}) \ ,
\end{equation}
where we have included the initial value of the field as the final argument.
Given that the field is linearly proportional to its initial value,
the above relation simplifies further to 
\begin{equation}
\label{relsyasy2}
\phi _{A}({\bf r},t;1) =  m_{\rm eq} - (1-m_{\rm eq})
\phi _{S}({\bf r},t;1) \ .
\end{equation}
Thus the average magnetization density relaxes to its equilibrium value with
the dynamics of the symmetric system (cf. Eq.(\ref{resden0})), 
up to a factor of $(1-m_{\rm eq})$.
Integrating this equation over space, and using the definition 
(\ref{glomagc1}) of the average global magnetization, it is easy to see that
the relation (\ref{mequ2}) is an exact consequence of this model.

We have tested the validity of (\ref{relsyasy2}) by measuring the average
magnetization density of a patch of spins in a numerical simulation, 
and have found reasonable agreement. 
We report on our numerical work in more detail in the next section.

\subsection{Quenched random rates}

We now consider an alternative generalization in which the rates are
symmetric, but spatially inhomogeneous. At the microscopic level, 
this is modeled by attaching to each lattice position ${\bf r}$
a quenched random variable $q_{\bf r}\in (0,1)$ drawn from some distribution 
$S(\lbrace q_{\bf r} \rbrace )$. The random variable $q_{\bf r}$ gives
the probability of switching in the event that the BA visits the site 
${\bf r}$. In the continuum theory, we model these quenched random couplings
by a simple extension of Eq.(\ref{scacont2}). We generalize
the coupling parameter
$\lambda $ to a spatially inhomogeneous random function $\lambda ({\bf r})$,
with a distribution $S[\lambda ]$. This does not affect the solvability of the
model due to the locality of the continuum theory, so that we have the
explicit solution
\begin{equation}
\label{solconrc}
\phi ({\bf r},t;\lambda) = \exp \left [ -\lambda ({\bf r}) \int \limits
_{0}^{t} dt' \ \Delta _{l} ({\bf r}-{\bf R}(t')) \right ] \ , 
\end{equation}
where we have emphasized the dependence of $\phi $ on $\lambda $.

There is an enormous variety in the forms of the quenched disorder distribution
$S$ that one may study. A comprehensive analysis is beyond the scope of the
present work. In our simulations (to be described in the next section) we
have chosen to study switching rates $q_{\bf r}\in (0,1)$ 
which are independent random variables drawn from a uniform distribution.
There is more than one way in which such a distribution can manifest itself
at the continuum level. Depending on the extent to which one coarse-grains
the lattice model, the continuum theory will have a distribution of
couplings which is either very similar to the lattice distribution ({\it i.e.}
uncorrelated and uniform), or more Gaussian in nature (due to the 
central limit theorem). We expect different physics according to the
behavior of the distribution $S[\lambda ]$ as $\lambda \rightarrow 0$.
Roughly speaking, if $S$ has zero weight for small-$\lambda $ we expect
the quenched disorder to be irrelevant to the system for large times.
Conversely, if $S$ has non-zero weight for small $\lambda $, then the
disorder will play a role for arbitrarily late times.
In order to exemplify this latter case, we consider 
as an example the distribution
\begin{equation}
\label{lamdist}
S[\lambda ] = \prod \limits _{\bf r} S_{\rm loc} \ (\lambda ({\bf r})) \ ,
\end{equation}
where $S_{\rm loc}$ is a local (or ``on-site'') distribution function which we
take to be uniform:
\begin{equation}
\label{lamdist2}
S_{\rm loc}(\lambda ) = (1/\lambda _{m}) \ H(\lambda _{m} - \lambda ) 
H(\lambda ) \ ,
\end{equation}
where $H(x)$ is the Heaviside step function\cite{jj}.
Henceforth we shall concentrate on $d=1$ for simplicity.

Due to the way in which the quenched random couplings enter the
continuum theory, we can easily generalize our earlier exact results.
Let us first concentrate on the magnetization density at the origin, 
averaged over both the BA trajectories, and the distribution of $\lambda $.
The first average is performed in the usual way using Laplace 
transforms (see DCI) and we have the analog of Eq.(\ref{ltden})
\begin{equation}
\label{ltdenra}
{\hat m}(0,s;\lambda) = {1 \over s} \left [ {1 
\over 1+\lambda (0) {\hat g}(0,s)} \right ] \ .
\end{equation}
On averaging this density over the distribution of random couplings
(\ref{lamdist2}) we have
\begin{equation}
\label{ltdenra2}
\langle {\hat m}(0,s;\lambda) \rangle _{S} = 
\left ( {2D \over \lambda _{m}^{2}s} \right )^{1/2} \log \left [ 1 
+ \left ( \lambda _{m}^{2} \over 2Ds \right )^{1/2} \right ] \ .
\end{equation}
Inverse Laplace transforming yields the asymptotic result
\begin{equation}
\label{ltdenra3}
\langle m(0,t;\lambda) \rangle _{S} = 
\left ( {D \over 2 \pi \lambda _{m}^{2}t} \right )^{1/2} \log 
\left ( \lambda _{m}^{2}t \over 2D \right ) \ .
\end{equation}
Therefore the decay in the homogeneous case (of $1/\sqrt{t}$ behavior)
for the average magnetization density is 
slowed by a logarithmic factor due to the presence of the quenched
random couplings.
It is interesting that this logarithmic slowing down does not affect
the average of the global magnetization. This quantity may be explicitly
calculated using the method of infinite 
order perturbation theory described in DCI.
The result is that the leading order term for large times is independent
of the coupling, and is unaffected by the average over $S$. 
Thus the average over the random couplings will in no way affect the
leading order behavior of $M(t) \sim (Dt)^{1/2}$. The average of the
global magnetization is a very robust quantity, and is unaffected by
scalings of the homogeneous couplings (as seen in section II) and by
making the couplings quenched random variables.
We have performed numerical simulations for systems with quenched random
couplings and report our findings toward the end of the following section.

Finally, we briefly discuss the changes that can occur to the probability
distribution ${\cal P}$ of the local magnetization density at the origin,
when quenched random couplings are introduced. We have from section II the
exact result (\ref{lognor}) for this function in $d=1$: it is a log-normal
distribution. The same form will hold in the present case, but now the
parameter $\lambda $ is to be averaged over $S[\lambda ]$. We
average over the uncorrelated uniform distribution given above in 
Eq.(\ref{lamdist2}). Simple integration gives
\begin{equation}
\label{lognorav}
\langle {\cal P} \rangle _{S} = {1 \over 2(\pi t)^{1/2} {\hat \lambda _{m}}
\phi } \ {\rm E}_{1} \left ( {(\log \phi )^{2} \over 4{\hat \lambda _{m}}^{2}
t } \right ) \ ,
\end{equation}
where ${\hat \lambda _{m}}=\lambda _{m}/(2D)^{1/2}$ and ${\rm E}_{1}(z)$
is the exponential integral\cite{as} . This distribution behaves very
much like the pure log-normal distribution for $\phi \ll 1$. However,
we have the interesting result that the distribution is singular at
$\phi = 1$. Explicitly we have for $\psi \equiv 1-\phi \ll 1$:
\begin{equation}
\label{lognorav2}
\langle {\cal P} \rangle _{S} \sim {\log [\psi ^{2}/4{\hat \lambda _{m}}^{2}
t] \over 2(\pi t)^{1/2} {\hat \lambda _{m}}} \ .
\end{equation}
This logarithmic
singularity stems from the fact that if $S[\lambda ]$ has a non-zero
weight for arbitrarily small $\lambda $, there will be a non-vanishing
subset of systems for which almost no corruption occurs. For large times,
this subset will appear as a pronounced ``peak'' in the distribution
$\langle {\cal P} \rangle _{S}$ for $\phi \rightarrow 1$.

\section{Numerical Simulations}

In this section we describe the details of our numerical simulations,
and present figures of our data to support the various theoretical
claims made in the preceding sections. The simulations are performed 
on a $d$-dimensional lattice (with $d$ either 1 or 2) and follow the 
lattice rules described in section II. We set the hopping probability
$p=1$, so that in each time step the BA is moved randomly to one of its
$2d$ nearest neighbor sites. In doing so, the spin it leaves behind
is flipped with probability $q$ (which is set to unity unless otherwise 
stated). Systems are taken large enough so that the BA never touches
the boundary during the lifetime of the simulation. We generally average
over between $10^{5}$ and $10^{7}$ realizations, depending on the desired
precision of the simulation. 

In section III we calculated the asymptotic form of the average 
magnetization density at the origin from
an exact lattice calculation. As noted, the sub-leading term is not
obtainable from the Tauberian theorem. However, it is clear from the
various calculations in section III that the corrections decay as
$1/(\log t)^{2}$. We have checked this in the following way. From
our simulations we measure $m({\bf 0},t)$ ( $ \ = \ \sum _{\bf R} 
\Theta ({\bf 0},{\bf R},n\delta t) \ $). We take its inverse and subtract the 
predicted asymptotic result of $(2/\pi)\log n$ (in order to compare with 
Eq.(\ref{finaldis2})). The resulting data is plotted against $1/\log n$, which
should yield a straight line, the $y$-intercept of which fixes $C(1)$.
As seen in the inset of Fig.2, 
the data indeed confirms this expectation, and we find $C(1) \simeq -0.742$. 
In Fig.2, we have plotted the inverse of the data against 
$\log t$, along with the predicted asymptotic result. The large difference
between the curves indicates the strong role of the corrections (which are
of order $10 \%$ even for $t \sim 10^{6}$).

We have also attempted to measure the probability distribution 
${\cal P}(\phi, {\bf 0}, t)$ in $d=2$, as calculated in section IIIE. In DCI
we attempted to measure this quantity in $d=1$ and we met with limited
success. The difficulty lies in the fact that the distribution only makes
sense for a coarse-grained magnetization density (since a single spin always
has a bimodal distribution). Thus, to construct ${\cal P}$ numerically 
we must measure the magnetization density for a patch of spins. If the patch
is too small, the resulting histogram will have too few bins to have any
smooth structure; but if the patch is too large, the time scales required
to allow the agent to wander far away from the patch and return many times
become numerically prohibitive. Thus we must compromise, and we use a square
patch
of 121 spins. The agent is started on the boundary of the patch to avoid
the patch being internally decimated by the transient motion of the agent.
The predicted result for ${\cal P}$ is given in Eq.(\ref{prob2}). 
It shows three
types of behavior set by a cross-over time $t^{*}$ defined by 
$\beta (t^{*})=1$. For $t \ll t^{*}$ the distribution has most weight near
$\phi =1$, and then for $t \sim t^{*}$ the distribution becomes uniform
over the entire interval of $\phi $. For late times $t \gg t^{*}$, the 
distribution approaches the form $\sim 1/\phi $. Our numerical
measurement of ${\cal P}$ is shown in Fig.3 for these three time regimes.
Good qualitative agreement is found. [Note that the measured distribution
has non-zero weight for negative $\phi $ due to the modest patch size,
and is also suppressed near $\phi =1$ due to internal decimation of
the patch magnetization.]

In section IV we presented an analysis of the continuum theory with $N$
agents. We described the interference between the agents by a number
termed the ``disordering efficacy'' $\sigma _{N}(d)$. In particular we
calculated the exact value for $\sigma _{N}(1)$ for $N=2,3$ and 4, as given
in Eq.(\ref{def5_1}). We also predicted the large-$N$ form of $\sigma _{N}(1)$,
along with the exact form of the strong logarithmic corrections, as shown
in Eq.(\ref{deflargen}). In two dimensions we argued that $\sigma _{N}(2)=N$ 
for all $N$, but only for times $t \gg e^{N}$. We have performed numerical 
simulations of the many agent system in order to test these results. The
microscopic rule we use is that there is no hard-core exclusion between
the agents, and that for each time step the $N$ agents are in
turn moved to a randomly chosen nearest neighbor site. A spin which is
occupied by two agents, say, will thus (for $q=1$) be flipped twice in 
that time step. In Fig.4 we show the evolution of the ratio of the average
global magnetization for $N$ agents as compared to one agent, for $d=1$. 
Asymptotically this ratio is the disordering efficacy by definition. Results
are shown for $N=2,3$ and 4. The curves asymptote to constants as expected, the
values of which are compared to the theoretical predictions. Excellent
agreement (to within less than $2\%$) is found. The numerical values are
also given in Table 1 in section IV. We have measured the disordering
efficacy for higher values of $N$ in $d=1$, and we plot our results on
a logarithmic scale in Fig.5. Also shown is the theoretical prediction
(\ref{deflargen}) for large-$N$ (where we have included both the leading
and sub-leading terms). Again, excellent agreement is found. As argued
in section IV, the cross-over times in the two-dimensional many agent system
are very large, growing exponentially with $N$. We have attempted to
measure $\sigma _{N}(2)$ numerically, but we have been unable to
reach large enough times to see the ratio of the average global magnetizations
reach a constant value. However, we may still compare our results to the
theoretical prediction, by plotting the predicted value of the asymptote
($=N$) minus the measured ratio, against $1/\log t$. If the predicted
asymptote is correct, the data so plotted should converge to the origin.
In Fig.6 we present such a plot for $N=2,3$ and 4. The resulting curves appear 
to be heading for the origin, and therefore constitute numerical support 
for the prediction $\sigma _{N}(2)=N$.

In the first part of section V we considered a system with asymmetric 
switching rates, and argued for the very simple phenomenological description
given by the equation of motion (\ref{contasy}). This equation may be
solved and one obtains the simple relationship (\ref{relsyasy2}) between 
the magnetization density for asymmetric rates and that for symmetric rates.
In particular, this result may be used to derive the intuitive result
(\ref{mequ2}) for the proportionality of the average global magnetization for
the asymmetric and symmetric cases ($M_{A}(t)$ and $M_{S}(t)$ respectively). 
We have performed simulations of the
system with asymmetric rates in $d=1$. In Fig.7 we show the ratio of
$M_{A}(t)/M_{S}(t)$ for different choices of the rates $q^{+}$ and $q^{-}$.
It is seen that the ratio gradually tends to a constant,
and that the constant is in agreement with the theoretical prediction
(\ref{mequ2}).
We have also tried to test Eq.(\ref{relsyasy2}) at the level of the average
magnetization density. We have found that the ratio of the densities is 
constant for all but the shortest times, which already indicates that the 
form of the time decay of $m(0,t)$ (relative to its equilibrium value) 
 is insensitive 
to the asymmetry in the rates. However, we have found that this constant is
sensitive to whether we measure the densities for an individual spin, or
for a patch. The value of the constant decreases as we increase the patch 
size. In Fig.8 we plot the ratio of magnetization densities for
asymmetric and symmetric systems, for various patch sizes with the rates
$q^{+}=0.8$ and $q^{-}=0.2$. As we see, the ratio seems to be approximately
constant, and the value decreases as the patch size is increased. Reasonable 
agreement is found with the prediction of Eq.(\ref{relsyasy2}) for the
largest patch of 101 spins. This indicates that (\ref{relsyasy2}) is 
likely to be correct, but only after a large degree of coarse-graining.

Finally we turn to the predictions of section VB, which concern the
system with symmetric quenched random rates. 
We investigated this situation numerically in $d=1$
by measuring the global magnetization, and the magnetization density.
We perform the averaging in batches; namely, we use the same set of quenched
rates for a batch of 1000 systems, which are then averaged over their 
different BA histories. Then we repeat this for $N_{b}$ 
batches (with $N_{b} \sim 10^{3}$), thus averaging over different realizations
of the quenched rates. We choose the quenched couplings $\lbrace q_{\bf r} 
\rbrace $ to be uncorrelated
random variables drawn from a uniform distribution in the range $[0,1]$.
In Fig.9 we show the average of the global magnetization in a 
system with quenched rates, and compare it to the same quantity in a ``pure''
system. The two curves become identical for late times (following a power
law growth $\sim \sqrt{t}$) indicating that the
average global magnetization is insensitive to the quenched rates, as 
predicted. Turning to the average magnetization density, we found at first
that our results could not be fitted to the theoretical prediction
(\ref{ltdenra3}). However, a closer analysis revealed that great care
must be taken in making the comparison. The point is that one wishes
to compare theory and simulation for long times (when the asymptotic 
form given in (\ref{ltdenra3}) becomes valid). However, it turns out that
there is a new cross-over time $t_{f}$ within the numerical simulations, beyond
which the theoretical prediction is expected to break down. This cross-over
time emerges due to insufficient averaging over batches. In the simulation
we average over $N_{b}$ batches, and thus select $N_{b}$ couplings from
the uniform distribution. There will be a smallest selected value 
$q_{f} \sim 1/N_{b}$ of the coupling,
and thus from this finite sampling of $S$, we cannot discern if we are really
using a uniform distribution with non-zero weight all the way down to $q=0$,
or a distribution with non-zero weight only down to $q=q_{f}$. In the latter
case, one can shown (from either the lattice or continuum theories) that
the logarithmic slowing down will vanish after a time $t_{f} \sim 1/q_{f}^{2}$.
Thus, in our simulations
we can only expect to see the logarithmic slowing down for times much less than
$t_{f} \sim N_{b}^{2}$. So, in order to makes this time-window as large as 
possible it is important to perform as many batch averages as possible,
at the expense of averaging over BA histories within a given batch.
To make a quantitative comparison to theory, we have calculated the exact
form of the logarithmic slowing down from the lattice theory, using the
methods described in section IIIA. After averaging over a uniform distribution
of the rates, we have from the Tauberian theorem the exact asymptotic result
\begin{equation}
\label{lattdis}
\left \langle \sum \limits _{R} \Theta (0,R,n\delta t) \right \rangle _{S}
\sim {\log (n) \over 2(2\pi n)^{1/2}} \left [ 1 + O \left ( 
{1 \over \log n} \right ) \right ] \ .
\end{equation}
We plot in Fig.10 the average magnetization density,
and the above asymptotic result. Good agreement is found, and the 
difference may be well fitted against the slow logarithmic corrections.
Note, for times longer than those shown on the plot, the data fails
to agree with the predicted result due to cross-over into the regime
$t \gg t_{f}$ in which finite sampling effects dominate.

\section{Summary and Conclusions}

This paper has been devoted to a continuation of our previous study DCI
\cite{nt}
of the statistical properties of data corruption due to a Brownian agent (BA).
In order to make closer contact to potential physical realizations of this
process we have extended our original study in three directions: i) a
careful examination of two dimensions, which is the critical dimension
for the model, ii) the competition induced by more than one agent, and iii)
the case of generalized couplings between the BA and its environment.

In section II we provided a recapitulation of DCI and outlined the discrete and
continuum modelizations of the process, along with presenting 
results pertinent to the present study. 

Section III was devoted to a careful examination of two dimensions, which, 
because of the recurrent properties of random walks, is the critical dimension
for this process. Thus, naive dimensional arguments are modified
by logarithmic corrections, and there are also strong corrections to the
leading order results. For these reasons it is difficult to obtain
controlled analytic results and to compare them meaningfully to numerical
data. In an attempt to overcome these obstacles we presented four
different methods of calculation for the average magnetization density at the 
origin $m({\bf 0},t)$, using
both discrete and continuum models. We found agreement in all cases for the
leading behavior $m({\bf 0},t) \sim 2\pi D/\lambda \log t$. We also gained
insight into the non-universal nature of the sub-leading terms, which in
all cases depends on the type of regularization used. However it is
clear from the various forms of $m({\bf 0},t)$ that the leading corrections
decay as $1/(\log t)^{2}$. This allowed us to analyze our data in a
quantitative
way, and excellent agreement was found with our numerical simulations 
for both the leading term, and for the
nature of the logarithmic corrections. Furthermore, given the explicit
functional forms derived for $m({\bf 0},t)$ we were able to derive 
approximate forms for the probability distribution ${\cal P}$ of the 
magnetization
density at the origin. The precise form is again non-universal as the
shape of ${\cal P}$ depends sensitively on the sub-leading terms of  
$m({\bf 0},t)$. However, we compared the simplest form (\ref{prob2})  
with simulation and found good qualitative agreement.

In section IV we generalized the model to a process with $N$ agents, which
although non-interacting, produce statistical interference in the corruption
of the system due to the overlap of their histories. We described their
effect via the ``disordering efficacy'' $\sigma _{N}(d)$ 
defined as the ratio of the 
average global magnetization due to $N$ agents, as compared to that of one
agent. After deriving a general form for $\sigma _{N}(d)$ (for $d<2$), we first
studied the case $d=1$. We were able to derive exact values for 
$\sigma _{N}(1)$ for $N=2,3$ and $4$ which are in excellent agreement with
simulations. Returning to general $d<2$ we found that 
$\sigma _{2}(d) = 2^{d/2}$, which assuming smooth continuation to two 
dimensions implies $\sigma _{2}(2)=2$, meaning that the two agents are 
statistically independent. We then formulated a large-$N$ analysis of the 
disordering efficacy, and found for $d<2$ that $\sigma _{N}(d) \sim 
[\log N]^{d/2}$. Thus in dimensions less than two, the agents interfere very
strongly with one another. The precise logarithmic dependence (with the
accompanying strong logarithmic corrections) are supported by excellent
quantitative agreement with simulations in $d=1$. We found that the 
large-$N$ analysis fails on continuation to $d=2$. An alternative analysis
for $d=2$ yielded the result $\sigma _{N}(2)=N$ which is satisfied
in the deep asymptotic regime $t \gg e^{N}$. This result is also supported
by numerical simulation in $d=2$ for modest values of $N$. Thus, the 
disordering efficacy is extremely sensitive to the dimensionality of the
system, and is strongly discontinuous as $d \nearrow 2$.

We moved on to a study of generalized couplings in section V. First
we considered homogeneous couplings (between the BA and the environment) which
are asymmetric. In the context of data corruption, this means that the
agent has a probability of switching $0 \rightarrow 1$ which is different
to the probability of switching $1 \rightarrow 0$. We proposed to model
this effect by a simple extension of the original continuum model, based on
the idea that the dynamics of the asymmetric process may be described 
relative to the non-zero background magnetization, using the relaxational
dynamics of the symmetric process. This idea leads to several simple
predictions for the average (local and global) magnetization which are
supported by simulation. Second, we considered symmetric but 
inhomogeneous couplings. Thus, the probability for the agent to switch
a given element is given by a quenched random variable. We incorporated
this new effect into the continuum model by generalizing the constant
parameter $\lambda $ to a quenched random field $\lambda ({\bf r})$ described
by a distribution $S[\lambda ]$. We investigated the simple case in which
$S$ is spatially uncorrelated, with a uniform on-site distribution function.  
We found that the average magnetization density at the origin decays
more slowly than in the homogeneous case, by a factor $\sim \log t$. However,
the average global magnetization is asymptotically insensitive to the
quenched couplings. The probability density for the 
magnetization density at the origin ${\cal P}(\phi ,0,t)$ is dramatically
altered for values of $\phi $ close to unity, where it suffers a 
logarithmic divergence; this is due to a non-vanishing subset of systems
for which the coupling $\lambda (x=0)$ is very small, such that barely
any corruption occurs. The main features of corruption in the presence
of quenched random couplings were supported by numerical simulation.

Finally in section VI we described the details of our numerical simulations,
and presented figures supporting the different analytic predictions from
the previous sections. 

In conclusion, we have presented three important extensions to our original
work. The most important fact to emerge from this investigation is the
robustness of the phenomenological model (\ref{scacont2}) in accounting
for the properties of the data corruption process under
a wide range of conditions. The fact that this process may be modeled
by such a simple continuum theory gives one confidence in applying
similar phenomenological models to more complicated physical situations
involving Brownian agents. 

As mentioned in the Introduction, the process of disordering of a bistable
medium by a Brownian agent has putative applications to data corruption,
homogeneous catalysis and the effect of impurities in biatomic crystals.
It remains to be seen whether one can find a solid application of the 
models studied in DCI and the present work. By having explored more challenging
situations such as two dimensions, many agents, and asymmetric and
quenched random couplings, we are able to seriously address the practical
issues involved in making the connection between the rich statistical
properties of our model, and their potential existence in the real world.

\newpage

The authors would like to thank B. Schmittmann for useful discussions.
The authors also gratefully acknowledge financial support from the 
Division of Materials Research of the National Science Foundation. 

\newpage

\appendix

\section{}

In this appendix we outline the procedure of taking the inverse Laplace 
transform of Eq.(\ref{ltcle2}). Consider the function $H(t)$, the Laplace
transform of which has the form 
\begin{equation}
\label{lt1}
{\hat H}(s) = {1 \over s[a - b\log s ]} \ , 
\end{equation}
for sufficiently small ${\rm Re} (s)$ 
such that the apparent pole at $s=e^{a/b}$ may
be disregarded ($a$ and $b$ are positive constants).
The inverse Laplace transform is given by the Bromwich integral\cite{jj}
\begin{equation}
\label{lt2}
H(t) = \int _{C} {ds \over 2\pi i} \ {e^{st} \over 
s[a - b\log s ]} \ , 
\end{equation}
where the contour $C$ is to be taken up the imaginary axis (note the
integration around the simple pole at the origin gives a contribution of zero).
The important singularity is the branch point at the origin. We take the
cut along the negative real axis and deform the contour around this cut.
Integrating across the cut then gives us 
\begin{equation}
\label{lt3}
H(t) = {1 \over b} \int \limits _{0}^{\infty} {dx \over x} \ 
\left [ {e^{-xt} \over (\log x - a/b )^{2} + \pi ^{2} } \right ] \ .
\end{equation}
We are interested in the asymptotic form of this integral for large $t$.
In this limit the integral is dominated by small $x$. We therefore expand
the denominator to give
\begin{equation}
\label{lt4}
H(t) = {1 \over b} \int \limits _{0}^{\infty} {dx \over x} \ 
{e^{-xt} \over (\log x)^{2}} \ \left [ 1 + {2a \over b\log x} + 
O\left ( {1 \over (\log x)^{2}} \right ) \right ] \ .
\end{equation}
By splitting the range of integration into two pieces ($x \in (0,1/t)$ and
$x \in (1/t, \infty)$ ) we systematically evaluate the terms in the 
above expansion to find
\begin{equation}
\label{lt5}
H(t) = {1 \over b\log t} - {(a + \gamma b)\over (b\log t)^{2}} +
O\left ( {1 \over (\log t)^{3}} \right ) \ ,
\end{equation}
where $\gamma = 0.57721...$ is Euler's constant. Using the appropriate form
for the constants $a$ and $b$  we arrive at Eq.(\ref{ltcle3}) correct up to 
$O(1/(\log t)^{3})$.

\section{}

In this appendix we indicate the steps leading from Eq.(\ref{def3}) to
Eq.(\ref{def4}). First we factorize the integrand to give
\begin{equation}
\label{def3a}
\sigma _{N}(d) = d\Gamma (1-d/2)\sum \limits _{n=0}^{N-1} J_{n}(d) \ ,
\end{equation}
where
\begin{equation}
\label{def3b}
J_{n}(d) = \int \limits _{0}^{\infty} dr \ r^{d-1} \
f(r) m(r)^{n} \ ,
\end{equation}
with $f(r) \equiv 1-m(r)$. We then integrate the above expression by parts
using 
\begin{equation}
\label{def3c}
\int \limits ^{r} du \ u^{d-1}f(u) = {r^{d}f(r) \over d} - {e^{-r^{2}} \over
d\Gamma (1-d/2)} + {\rm constant} \ . 
\end{equation}
The resulting integral contains a factor of $re^{-r^{2}}$ which allows
one to integrate by parts a second time, thus yielding after some
algebra (for $n>0$)
\begin{equation}
\label{def3d}
J_{n}(d) = -{\delta _{n,1} \over d\Gamma (1-d/2)} + K_{n}(d) - K_{n-1}(d) \ ,
\end{equation}
with 
\begin{equation}
\label{def3e}
K_{n}(d) = {2n(n+1) \over d\Gamma (1-d/2)^{2}} \int \limits _{0}^{\infty}
dr \ r^{1-d} \ e^{-2r^{2}} \ m(r)^{n-1} \ .
\end{equation}
Substitution of Eq.(\ref{def3d}) into (\ref{def3a}) yields the result
(\ref{def4}) as shown in the main text.

\section{}

In this appendix we indicate the evaluation of $\sigma _{N}(1)$ as given
by Eq.(\ref{def4_1}) for $N=3$ and $N=4$. For $N=3$ we require
\begin{equation}
\label{def4a}
\sigma _{3}(1) = {12 \over \pi ^{1/2}} \int \limits _{0}^{\infty} dx \
e^{-2x^{2}} \ {\rm erf}(x) \ .
\end{equation}
We introduce the integral
\begin{equation}
\label{def4b}
I(\beta ) \equiv \pi ^{1/2}\int \limits _{0}^{\infty} dx \ e^{-\beta x^{2}} \ 
{\rm erf}(x) \ .
\end{equation}
Using integration by parts it is straightforward to show that $I(\beta )$
satisfies the differential equation
\begin{equation}
\label{def4c}
2\beta {dI \over d\beta} + I = -{1 \over (1+\beta )} \ .
\end{equation}
This differential equation may then be integrated (using the boundary 
condition $I(\infty)=0$) to give
\begin{equation}
\label{def4d}
I(\beta ) = {1 \over 2\beta ^{1/2}} \int \limits _{\beta }^{\infty}
d\beta ' {1 \over \beta '^{1/2}(1+\beta ')} = {1\over \beta ^{1/2}}
{\rm sin}^{-1} \left ( {1 \over (1+\beta)^{1/2}} \right ) \ .
\end{equation}
Using this result along with the fact that $\sigma _{3}(1) = (12/\pi)I(2)$, 
we have the leftmost result of Eq.(\ref{def5_1}) as shown in the main text.

The evaluation of $\sigma _{4}(1)$ proceeds along similar grounds. We define
a function
\begin{equation}
\label{def4e}
L(\beta ) \equiv \pi ^{1/2}\int \limits _{0}^{\infty} dx \ 
e^{-\beta x^{2}} \ {\rm erf}(x)^{2} \ ,
\end{equation}
which may be shown to satisfy the differential equation
\begin{equation}
\label{def4f}
2\beta {dL \over d\beta} + L = -{2 \over (1+\beta )(2+\beta )^{1/2}} \ .
\end{equation}
This equation may be integrated, and the resulting integral may be calculated
by elementary methods to yield
\begin{equation}
\label{def4g}
L(\beta ) = {1 \over \beta ^{1/2}} {\rm sin} ^{-1} \left ( 
1\over 1+\beta \right ) \ .
\end{equation}
Using this result along with the fact that $\sigma _{4}(1) = (24/\pi)L(2)$, 
we have the rightmost result of Eq.(\ref{def5_1}) as shown in the main text.

A similar method may be used to simplify the integrals for higher $N$,
but we have been unable to evaluate these integrals in closed form. Naturally,
the integral in Eq.(\ref{def4_1}) may be evaluated numerically for a given
$N$ to any desired precision.

\begin{figure}[tbp]
\centerline{\epsfxsize=13.0cm
\epsfbox{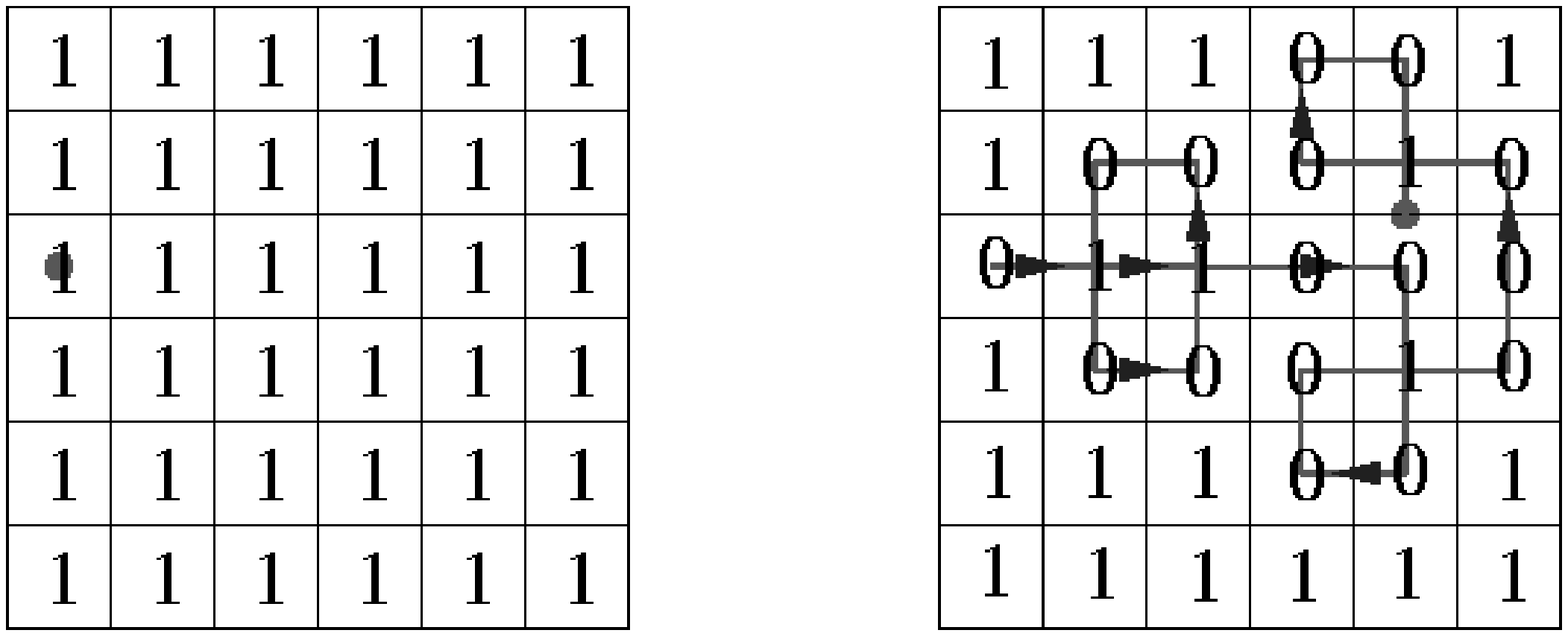}}
\vspace{0.1cm}
\caption{Illustration of the data corruption process for $d=2$,
with $p=q=1$. The
initial uncorrupted state is shown on the left, with the BA represented
by the filled circle. On the right we show a typical walk of $\sim 20$ steps.
The BA switches a bit with each visit, so those bits visited an even number
of times are restored to their original value.}
\end{figure}

\begin{figure}[tbp]
\centerline{\epsfxsize=10.0cm
\epsfbox{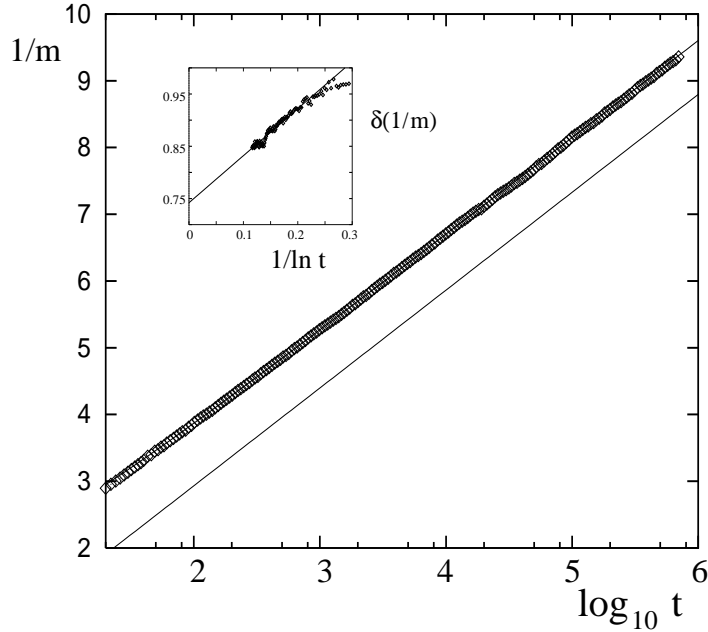}}
\vspace{0.1cm}
\caption{Plot of $1/m({\bf 0},t)$ 
(the inverse of the average magnetization density at the
origin), as a function of $\log t$ for $d=2$. 
The lower solid line is the asymptotic
prediction, while the upper solid line (which is partially obscured by
the data) is the prediction with corrections
to scaling included. The inset shows the difference between $1/m({\bf 0},t)$ 
and its asymptotic form, plotted against $1/\log t$. The data may be
fitted to a straight line, thereby yielding the corrections to scaling
explicitly.}
\end{figure}

\begin{figure}[tbp]
\centerline{\epsfxsize=10.0cm
\epsfbox{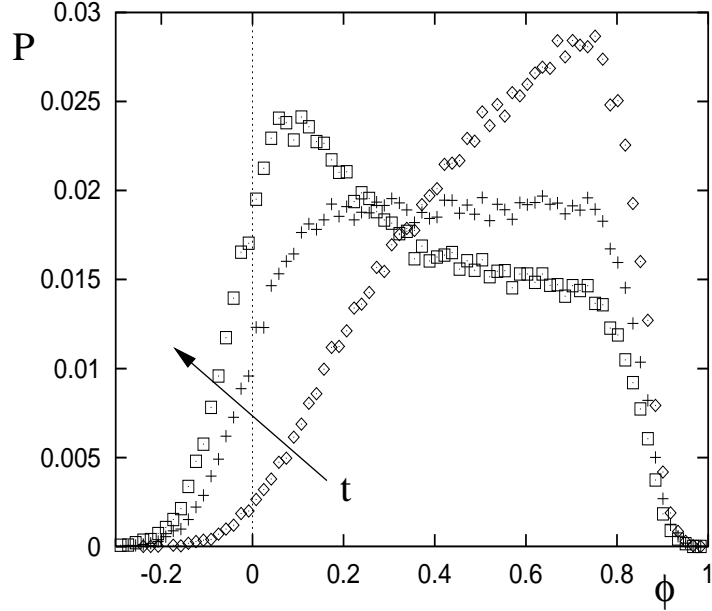}}
\vspace{0.3cm}
\caption{Plot of the probability distribution ${\cal P}$ as a function
of $\phi $ for $d=2$, 
for times $t=10^{3}$ (diamonds), $t=10^{4}$ (plusses), and
$t=10^{5}$ (squares). Note the qualitative similarity (as time proceeds)
between these
different histograms, and the theoretical prediction (\ref{prob2}).}
\end{figure}

\begin{figure}[tbp]
\centerline{\epsfxsize=10.0cm
\epsfbox{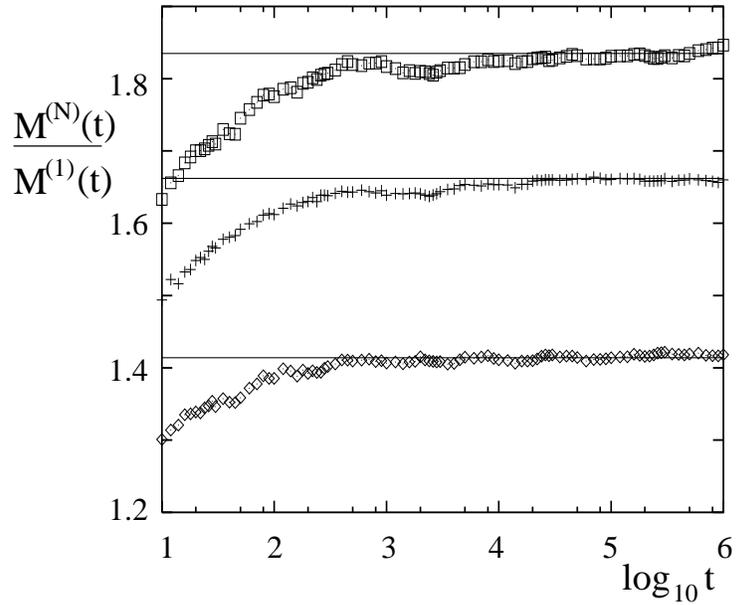}}
\vspace{0.3cm}
\caption{Plot of the ratio of the average global magnetization for
$N$ agents as compared to a single agent, for (in ascending order) $N=2,3,$ 
and 4, in $d=1$. 
The solid lines are the theoretical predictions (\ref{def5_1}) for the 
disordering efficacies, which are the asymptotes of this ratio.}
\end{figure}

\begin{figure}[tbp]
\centerline{\epsfxsize=10.0cm
\epsfbox{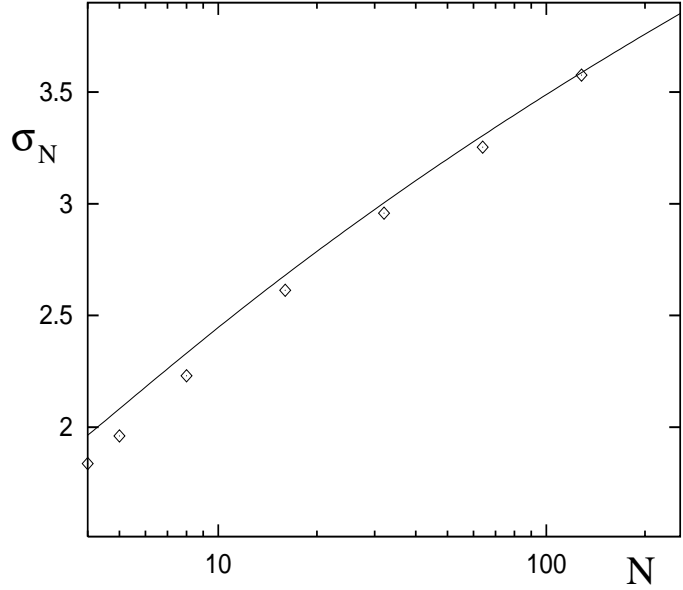}}
\vspace{0.3cm}
\caption{Plot of the disordering efficacy $\sigma _{N}$ as a function
of $N$ in $d=1$. The solid line is the theoretical prediction
(\ref{deflargen}) (including the strong logarithmic corrections).}
\end{figure}

\begin{figure}[tbp]
\centerline{\epsfxsize=10.0cm
\epsfbox{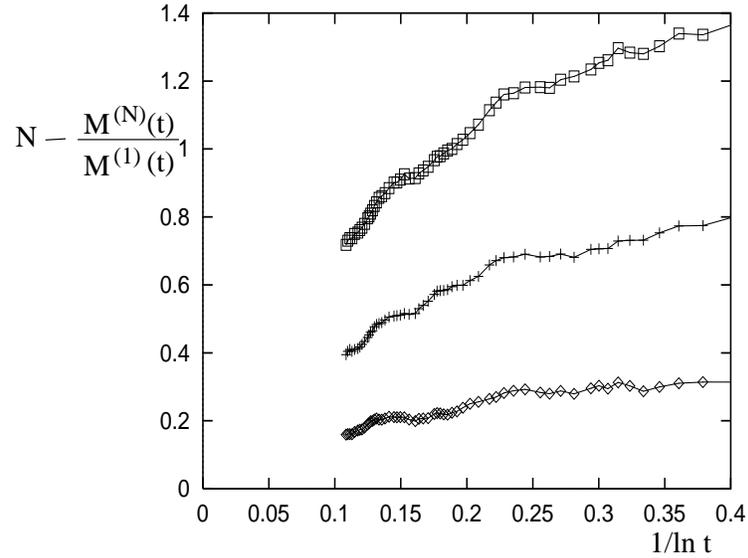}}
\vspace{0.3cm}
\caption{Plot of the ratio of the average global magnetization for
$N$ agents as compared to a single agent, subtracted from $N$, as a 
function of $1/\log t$ for 
(in ascending order) $N=2,3,$ and 4, in $d=2$. The data in each case
is apparently heading for the origin, thus supporting the theoretical
prediction $\sigma _{N}=N$.}
\end{figure}

\begin{figure}[tbp]
\centerline{\epsfxsize=10.0cm
\epsfbox{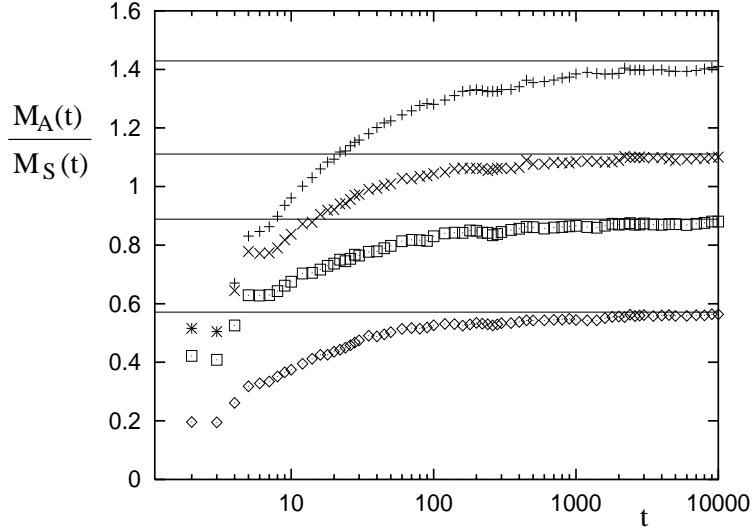}}
\vspace{0.3cm}
\caption{Plot of the ratio $M_{A}(t)/M_{S}(t)$ versus time for 
systems in $d=1$ with asymmetric rates. In ascending order, the
values of the rates are $(q^{-},q^{+}) = (0.5,0.2) , \ (0.5,0.4) , \ 
(0.4,0.5) , \ (0.2,0.5)$. The solid lines are the theoretical predictions
obtained from (\ref{mequ2}).}
\end{figure}

\begin{figure}[tbp]
\centerline{\epsfxsize=10.0cm
\epsfbox{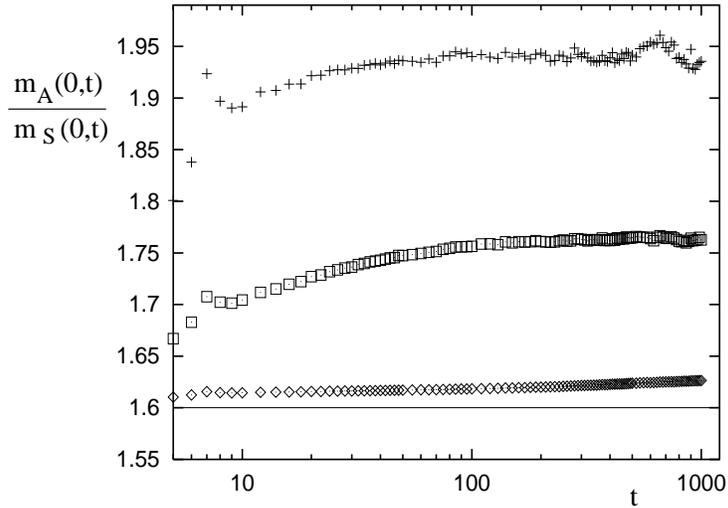}}
\vspace{0.3cm}
\caption{Plot of the ratio of magnetization densities $m_{A}(0,t)/m_{S}(0,t)$
for $d=1$ with asymmetric rates $q^{-}=0.2$ and $q^{+}=0.8$. The ratio is
measured for different patch sizes, the patches having from top to bottom
9, 19, and 101 spins respectively. The solid line is the theoretical
prediction from Eq.(\ref{relsyasy2}).}
\end{figure}

\begin{figure}[tbp]
\centerline{\epsfxsize=10.0cm
\epsfbox{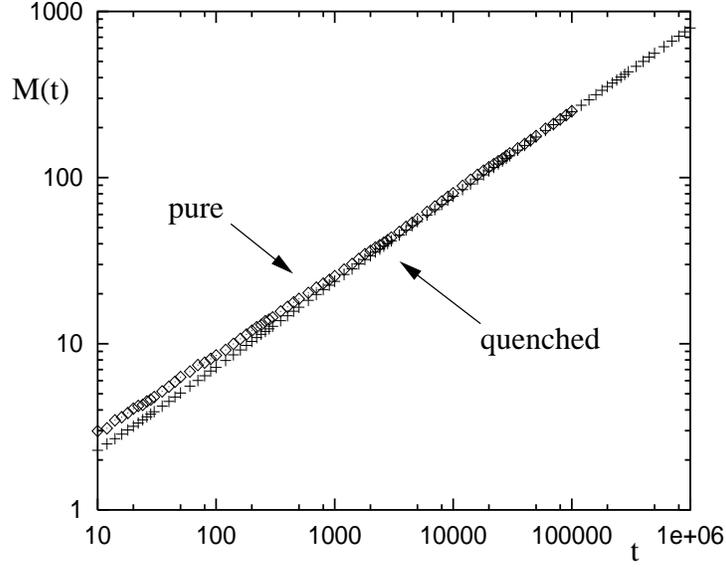}}
\vspace{0.3cm}
\caption{Plot of the average global magnetization versus time 
for systems with quenched and pure (homogeneous) switching rates. For
long times the functions become identical, growing $\sim \sqrt{t}$.}
\end{figure}

\begin{figure}[tbp]
\centerline{\epsfxsize=10.0cm
\epsfbox{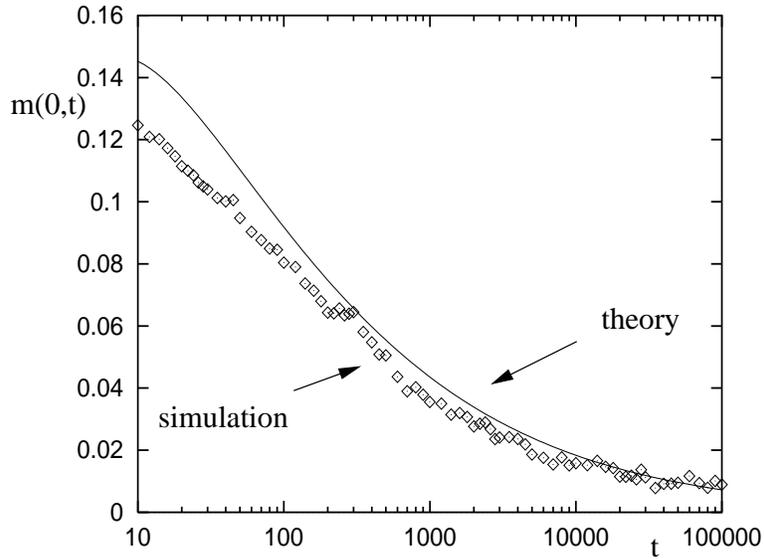}}
\vspace{0.3cm}
\caption{Plot of the magnetization density at the origin
(averaged over both BA histories and the quenched rates) versus
time. The solid line is the asymptotic form of the 
theoretical prediction Eq.(\ref{lattdis}) with no free parameters.}
\end{figure}
\end{document}